\pdfoutput=1
\documentclass[11pt,letterpaper]{article}
\usepackage[margin=1in]{geometry}
\usepackage[T1]{fontenc}
\usepackage{lmodern,microtype}
\usepackage{amsmath,amssymb,amsthm,mathtools,bm}
\usepackage{booktabs,array,xcolor,graphicx}
\usepackage{flafter}
\usepackage[numbers,sort&compress]{natbib}
\usepackage[colorlinks=true,linkcolor=black,citecolor=black,urlcolor=blue]{hyperref}
\newtheorem{theorem}{Theorem}
\newtheorem{lemma}[theorem]{Lemma}
\newtheorem{corollary}[theorem]{Corollary}
\newtheorem{proposition}[theorem]{Proposition}
\newtheorem{claim}[theorem]{Claim}
\theoremstyle{definition}\newtheorem{definition}[theorem]{Definition}

\usepackage{etoolbox}
\AtBeginEnvironment{theorem}{\samepage}
\AtBeginEnvironment{lemma}{\samepage}
\AtBeginEnvironment{corollary}{\samepage}
\AtBeginEnvironment{proposition}{\samepage}
\AtBeginEnvironment{claim}{\samepage}
\AtBeginEnvironment{definition}{\samepage}
\newcommand{\needlines}[1]{\par\ifdim\pagetotal>0pt\ifdim\dimexpr\pagegoal-\pagetotal\relax<#1\baselineskip\penalty-10000\fi\fi}
\newcommand{\C}{\mathbb C}
\newcommand{\E}{\mathbb E}

\newcommand{\Tr}{\operatorname{Tr}}
\newcommand{\TD}{\operatorname{TD}}
\newcommand{\Ad}{\operatorname{Ad}}
\newcommand{\Dom}{\operatorname{Dom}}
\newcommand{\Img}{\operatorname{Im}}
\newcommand{\ket}[1]{\lvert#1\rangle}
\newcommand{\bra}[1]{\langle#1\rvert}
\newcommand{\proj}[1]{\ket{#1}\!\bra{#1}}
\newcommand{\vket}[1]{\lvert#1\rangle\!\rangle}
\newcommand{\vbra}[1]{\langle\!\langle#1\rvert}
\newcommand{\norm}[1]{\lVert#1\rVert}
\newcommand{\SW}{\mathcal S}
\newcommand{\QC}{\mathcal Q}
\newcommand{\id}{\operatorname{id}}

\newcommand{\reffig}[1]{Figure~\ref{#1}}
\newcommand{\before}{\prec}

\hypersetup{pdftitle={Infinite or indefinite: the query cost of simulating the quantum switch},pdfauthor={Dantong Li}}
\title{Infinite or indefinite:\\ the query cost of simulating the quantum switch}
\author{Dantong Li\\\small Department of Computer Science, Yale University}
\date{November 2026}
\usepackage{needspace}
\begin{document}
\maketitle

\begin{abstract}
The quantum switch applies two operations in a coherent superposition of their causal orders and is a simple example of indefinite causal order. No quantum circuit that uses each operation once in a definite order can even describe it, yet photonic experiments have demonstrated its physical realization. We ask how many calls to the black-box operations a quantum computer in the circuit model needs to reproduce the switch. If one of the black boxes is known to be unitary, three calls suffice. For general channels on a $d$-dimensional system, we prove three results. In all three, circuits may use classical feed-forward and unlimited ancillary quantum memory. First, no finite number of calls suffices for exact simulation in the worst case. Second, approximation to any fixed error below $1/2$, measured in half the diamond norm, requires $\Omega(d^{1/4})$ calls, exponential in the number of qubits. Third, an ordinary circuit can require exponentially many calls to obtain a single bit that the switch extracts with a few uses. We consider a device whose actual noise is dephasing in a fixed unknown basis and test it against a theoretical depolarizing model. The two channels have the same entanglement and average gate fidelities, so these figures alone do not distinguish them. At fixed noise strength, the quantum switch tells them apart with a number of device uses independent of $d$, whereas the circuit model requires $\Omega(d^{1/16})$ calls. Noise characterization is needed for quantum error correction, and this gives a provable query advantage of the switch in a task relevant to practical quantum computing. We invite future work that extends such advantages to more practical computing tasks.
\end{abstract}

\section{Introduction}

\begin{figure}[t]
\centering
\includegraphics{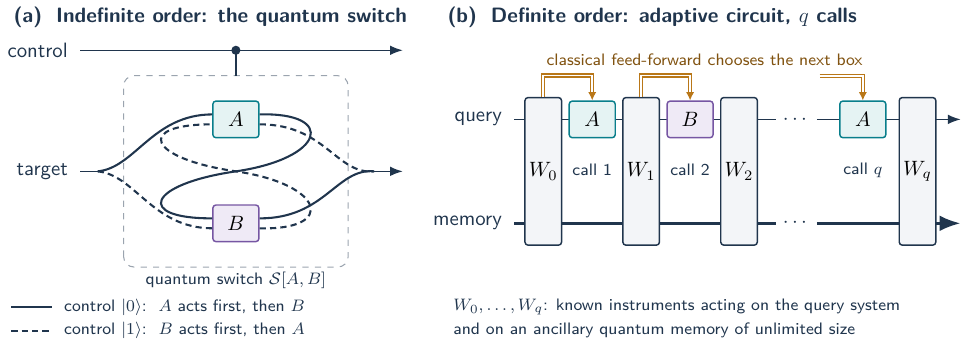}
\caption{The two kinds of access compared in this paper. (a)~The quantum switch: a control in superposition puts the two orders in superposition, and each operation is used once. (b)~An ordinary strategy: an adaptive circuit in which classical feed-forward from earlier outcomes decides which black box is called next, so the order of calls is definite on each run; the sequence $A,B,\dots,A$ is one possible run.}
\label{fig:switch}
\end{figure}

In a quantum circuit, operations are applied one after another. Quantum theory also allows processes in which the order of two operations is itself a quantum degree of freedom. The simplest is the \emph{quantum switch}~\cite{chiribella2013}. It takes two operations $A$ and $B$ and a control qubit. If the control is in the state $\ket0$, a target system undergoes $A$ and then $B$. If the control is in the state $\ket1$, the target undergoes $B$ and then $A$. If the control is in a superposition, so is the order, as illustrated in \reffig{fig:switch}(a). Each operation is used exactly once: the output depends only on the two operations and is linear in each of them. No quantum circuit that uses each operation once in a definite order does this~\cite{chiribella2013,araujo2015}. The switch is nevertheless physically realizable: photonic experiments in several laboratories have demonstrated it~\cite{procopio2015,rozema2024}. It has been reported to help in discriminating channels~\cite{chiribella2012}, in promise problems on unknown gates~\cite{araujo2014}, in communication~\cite{guerin2016,ebler2018}, in metrology~\cite{zhao2020}, and in measuring the incompatibility of quantum measurements~\cite{gao2023}. Some of these advantages have since been reassessed against other resources, such as coherent control of the channels~\cite{abbott2020} and more general strategies with a definite order~\cite{mothe2024}.

How much is an indefinite order worth as a resource? One way to put a price on it is to ask what an ordinary quantum computer must pay to do without it. We give the computer two \emph{black~boxes}, devices that apply unknown operations $A$ and $B$ when they are called. An \emph{ordinary strategy} is an adaptive quantum circuit that calls the black boxes one at a time, as in \reffig{fig:switch}(b). It has unlimited ancillary quantum memory and may apply any known operations between calls. It has classical feed-forward: it decides which box to call next from what it has measured so far. The order of its calls is therefore definite on every run, although it may differ from run to run. All our results hold against such strategies. We ask:\nopagebreak
\begin{quote}
\emph{How many calls to the black boxes $A$ and $B$ does an ordinary strategy need in order to reproduce what the quantum switch does with one call to each?}
\end{quote}

If one of the two black boxes is known to be unitary, three calls suffice: a circuit calls the unitary box twice and the other box once~\cite{chiribella2013,bavaresco2025}. The black boxes met in practice are rarely unitary, however. A noisy gate, a measurement whose outcome is not read, and a process coupled to an environment are described by quantum \emph{channels}. For general channels, the question is much harder. Bavaresco et al.~\cite{bavaresco2025} proved that no quantum circuit simulates the switch on two $n$-qubit channels with one call to one of them and fewer than $2^n$ calls to the other.

This paper answers our question at three levels of demand: reproducing the switch exactly, reproducing it approximately, and obtaining a single bit that the switch extracts. For the first, no finite number of calls suffices. The other two cost a number of calls that is exponential in the number of qubits. The third also gives a provable query advantage of the quantum switch in noise model characterization, a task relevant to quantum computing.

Earlier query advantages depend on the oracle model. For promise problems on unknown unitary gates, Renner and Brukner improved the known causal algorithms~\cite{renner2021} and proved advantages for a superposition of orders of qubit gates~\cite{renner2022}. For standard Boolean oracles, quantum control of causal order does not reduce query complexity~\cite{abbott2024}, although more general causally indefinite processes can~\cite{abbott2025}. Bavaresco, Murao, and Quintino~\cite{bavaresco2021} showed that strategies with indefinite order can improve the probability of discriminating two specified channels at a fixed number of calls. Our results concern the cost of simulating the two-channel switch and distinguishing noise models, with nonunitary channels supplying the hard instances.

\begin{figure}[t]
\centering
\includegraphics{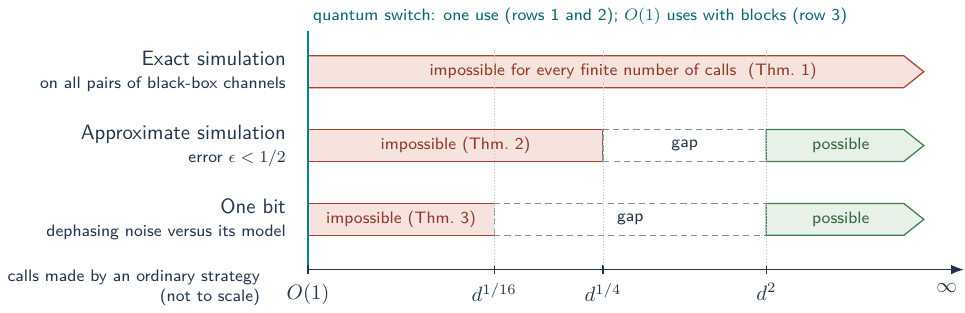}
\caption{Query costs for channels of dimension $d=2^n$, at fixed error and, in the third row, fixed noise strength. Red regions are excluded by Theorems~\ref{thm:exact-simulation}, \ref{thm:simulation-summary}, and~\ref{thm:c6}. Green regions are achievable uniformly: for every pair of channels in the second row and every channel in the promised family in the third. The constructions appear in Sections~\ref{sec:simulation} and~\ref{sec:c6-proof}. Dashed regions are the gaps between the bounds, and constants are omitted. For the third task, each switch slot contains a block of consecutive device calls.}
\label{fig:landscape}
\end{figure}

\subsection{Our results}\label{sec:results}

Throughout, the black boxes $A$ and $B$ are unknown channels on a $d$-dimensional system, and $\SW[A,B]$ is the switch, a channel on the control and the target. Theorems~\ref{thm:exact-simulation} and~\ref{thm:simulation-summary} hold for every $d\ge2$; Theorem~\ref{thm:c6} is stated for $n$ qubits, $d=2^n$. An ordinary strategy has \emph{budget} $q$ if it makes at most $q$ calls in total on every run. The least budget with which a task can be performed is its worst-case query complexity; it is infinite if no budget suffices. Section~\ref{sec:framework} gives the definitions, and \reffig{fig:landscape} summarizes the results.

\Needspace{9\baselineskip}
\paragraph{Exact simulation has infinite query complexity.}
\begin{theorem}[Exact simulation]\label{thm:exact-simulation}
For every $d\ge2$ and every finite budget $q$, no ordinary strategy with budget $q$ simulates the switch exactly on $\C^d$, that is, implements $\SW[A,B]$ for all channels $A,B$ on $\C^d$. This holds even if the strategy is required to work only for the control input $\ket+$. The target input is arbitrary, and the output consists of the control and the target.
\end{theorem}

The theorem proves Conjecture~1 of Bavaresco et al.~\cite{bavaresco2025} and answers their open question of whether any finite number of calls suffices: the worst-case query complexity of exact simulation is infinite in every dimension. The simulator is chosen without knowing $A$ and $B$ and must work for every pair of channels. It may be the full adaptive circuit in \reffig{fig:switch}(b), with classical feed-forward and quantum memory. This includes every circuit with classical control of causal order~\cite{wechs2021} and, by the recent result of Wechs et al.~\cite{wechs2026}, every causally separable process. The statement is a worst-case one: restricted classes of pairs can be easy, including pairs in which one box is known to be unitary. We next ask whether allowing a small error makes universal simulation efficient.

\paragraph{Approximate simulation is expensive.}
The error $\operatorname{err}(\QC;d)$ of a strategy $\QC$ is its largest distance to the switch over all pairs of boxes, measured in half the diamond norm as defined in Eq.~\eqref{eq:simulation-error}.

\begin{theorem}[Approximate simulation]\label{thm:simulation-summary}
For every fixed $0\le\epsilon<1/2$, an ordinary strategy that simulates the switch on $\C^d$ with error at most $\epsilon$ requires $\Omega_\epsilon(d^{1/4})$ calls. The lower bound holds already when the boxes are restricted to uniform mixtures of unitary channels.
\end{theorem}

The threshold $1/2$ is natural: measuring the control and then calling the boxes in the corresponding order gives error at most $1/2$ with two calls. In the other direction, $O(d^2/\epsilon)$ calls suffice for any positive error $\epsilon$, as shown in Section~\ref{sec:simulation}. Thus, for fixed positive error below $1/2$, the cost lies between $d^{1/4}$ and $d^2$ and is exponential in the number of qubits.

\paragraph{A single bit from the switch can already be hard.}
Simulation asks for the whole output of the switch on every input. Gao et al.~\cite{gao2023} used a measurement of the switch's control qubit to quantify the incompatibility of quantum measurements. We show that even a single bit obtained from this measurement can require exponentially many ordinary calls. Related measurements of control interference appear in metrology, communication, refrigeration, and the estimation of unitary invariants~\cite{ebler2018,zhao2020,felce2020,azado2025}.

Our example asks whether a depolarizing noise model describes a device. Depolarizing noise is a standard model in analyses of quantum error correction~\cite{wootton2012}. We consider a device whose actual noise is dephasing of known strength $0<p\le1$ in an unknown basis, and compare it with a depolarizing model of the same fidelity:
\[
\begin{aligned}
 \text{actual noise}\qquad
 &\mathcal E_{p,U}=(1-p)\id+p\,\Delta_U,\\
 \text{theoretical model}\qquad
 &\mathcal M_p=(1-p)\id+p\,\Lambda,
 \qquad \Lambda=\E_U\Delta_U.
\end{aligned}
\]
Here $\Delta_U$ dephases in the basis formed by the columns of $U$, which stays fixed at every use. The channel $\Lambda$ averages this dephasing over all bases and is depolarizing. The actual noise and its model have the same entanglement and average gate fidelities, so these figures alone do not tell them apart.

\begin{theorem}[One bit: does the depolarizing model describe the noise?]\label{thm:c6}
Let $d=2^n$, $n\ge1$, and $0<p\le1$. A device applies the same channel $\mathcal E$ at every use. It is promised that either the depolarizing model describes the device, $\mathcal E=\mathcal M_p$, or its noise is dephasing in an unknown basis, $\mathcal E=\mathcal E_{p,U}$ for some unitary $U$. Every ordinary strategy that decides between these cases with success probability at least $2/3$ for $\mathcal M_p$ and for every $\mathcal E_{p,U}$ requires $\Omega(d^{1/16}/p)$ calls. A constant number of uses of the switch, with repeated device calls in each slot, decide the question with one-sided error below $1/3$ using $O(1/p)$ device calls in total.
\end{theorem}

At fixed noise strength, the switch therefore answers the question with a number of calls independent of the dimension, while an ordinary strategy needs exponentially many in the number of qubits. Section~\ref{sec:c6-proof} gives the test and explains which feature of the noise it detects.

\subsection{Technical overview}\label{sec:overview}

\paragraph{Exact simulation.}
We argue by contradiction. Suppose a circuit with finite budget $q\ge1$ simulates the switch for every pair of channels. We first pad it to make exactly $q$ calls to each box. It must also work when each channel is mixed with the identity. Varying the two mixing probabilities gives an equality of polynomials. Comparing the terms in which exactly one call to each nonidentity channel remains shows that we can keep one randomly chosen call to each box and replace all other calls by the identity. Lemma~\ref{lem:polarization} shows that this new circuit produces a \emph{diluted switch}: the switch has weight $1/q^2$, and the remaining terms use at most one nonidentity channel.

\looseness=-1 The usual one-call impossibility result does not yet give a contradiction, because the new circuit only needs to reproduce this mixture. We therefore prove that even this diluted switch cannot be implemented with one call to each box, however small its positive switch weight is. For qubits, the Choi operator of such a circuit splits into two positive operators, one for each order, as described in Section~\ref{sec:ordered}. Positivity restricts both operators to a common subspace. Within that subspace, every definite-order circuit obeys a linear identity that the diluted switch violates. Section~\ref{sec:exact} and Appendix~\ref{sec:certificate} prove this identity and extend the contradiction from qubits to every dimension. The finite query budget matters because it leaves a strictly positive weight of the switch after the reduction.

\paragraph{Approximate simulation.}
The three-call construction for a unitary input~\cite{bavaresco2025} shows that the hard instances must involve nonunitary channels. We construct them as uniform mixtures of $r$ Haar-random unitaries in each box. Every call selects an index independently. We compare this with a process that samples indices without replacement. Their outputs are close when $q^2\ll r$, because the probability of a repeated index is small. In the new process, the circuit may call each channel many times, but it uses each selected unitary at most once. Fixing the other unitaries leaves a circuit with at most one call to each member of a chosen pair. Its output splits into the two possible query orders.

A circuit that calls $U$ before $V$ can apply $VU$ directly. We bound how well it can reproduce the reversed product $UV$. For a maximally entangled input, Lemma~\ref{lem:reverse} bounds its average probability of passing a test for $UV$ by $O(d^{-2})$. This limits the coherence it can produce between the two orders to order $1/d$, while the switch preserves full coherence. We combine these bounds over the $r^2$ pairs of unitaries. When $r^2\ll d$, a single measurement distinguishes the two outputs with constant bias. Section~\ref{sec:approximation} gives this measurement and accounts for the probability of a repeated index.

\paragraph{One bit.}
Here the difficulty is to detect a fixed preferred basis when that basis is unknown. An ordinary strategy can keep quantum memory and adapt later queries to earlier outcomes, so it is not enough to analyze independent probes. We compare the final states of the whole experiment under the two hypotheses, averaging over a basis chosen once and held fixed throughout.

Dephasing in an unknown basis can be written as a rotation into that basis, dephasing, and a rotation back. The path-recording framework of Ma and Huang~\cite{mahuang2025} keeps track of what queries to an unknown Haar-random unitary can reveal. We use it to compare both experiments with a simpler process whose stored basis labels are invariant under a common rotation. This symmetry prevents the labels from favoring a basis. The key estimate shows that dephasing one of a small number of stored systems is close to depolarizing it. Replacing the calls one at a time then shows that the two final states stay close when the query budget is small. This bounds the information available for the decision itself, even with quantum memory and adaptivity. We first prove the comparison at full noise strength, then reduce general $p$ to that case in Section~\ref{sec:c6-proof}.

\section{Background}\label{sec:framework}

\subsection{Channels and the switch}

A black box applies a fixed quantum channel, a completely positive trace-preserving map on operators on $\C^d$. Each call acts on a chosen $d$-dimensional system, which may be entangled with other registers. The box is memoryless and supplies only its quantum output: we have no direct access to its environment or to a separate classical record of its internal measurement outcomes.

For channels $A(\rho)=\sum_iA_i\rho A_i^\dagger$ and $B(\rho)=\sum_jB_j\rho B_j^\dagger$, the quantum switch is the channel on a control qubit and a target with Kraus operators
\begin{equation}\label{eq:switch}
 K_{ij}=\proj0\otimes B_jA_i+\proj1\otimes A_iB_j.
\end{equation}
The resulting channel $\SW[A,B]$ is independent of the chosen Kraus representations and is linear in each input channel~\cite{chiribella2013}. The same pair $A_i,B_j$ occurs in both orders, so a superposition of the control states preserves interference between $B_jA_i$ and $A_iB_j$.

\subsection{Ordinary strategies}\label{sec:model}

We compare the switch with quantum circuits whose next query can depend on earlier measurement outcomes. This is classical control of causal order in the sense of Wechs et al.~\cite{wechs2021}, with repeated calls to each box.

\begin{definition}[Ordinary strategy]\label{def:ordinary}
An \emph{ordinary strategy}, or \emph{adaptive circuit}, with budget $q$ has an input system, a $d$-dimensional query system, quantum memory of arbitrary finite dimension, and known instruments $W_0,\ldots,W_q$ acting on the query and memory. The input may be trivial. The instrument $W_0$ receives it, and each $W_t$ may depend on all earlier outcomes. After $W_t$, the recorded outcomes determine whether to return an output or make another call, and which box to call. Every run makes at most $q$ calls in total. Summing over all outcome sequences gives the implemented channel $\QC[A,B]$. The definition extends to any finite set of boxes.
\end{definition}

Each $W_t$ may contain measurements and conditional operations, so classical feed-forward is allowed both within an instrument and between instruments. Ancilla preparation, reset, and coherent control of known operations are included. The choice between unknown boxes is classical. There is no uniform bound on the memory dimension. For full simulation, the input and output contain the control and target; for the noise-model decision, the output is a classical bit.

We may add calls on dummy systems and discard their outputs without changing the implemented channel. In particular, a circuit with budget $q$ can be \emph{padded} to call each of two boxes exactly $q$ times. We use this in the exact proof.

\subsection{Simulation error and visibility}\label{sec:visibility}

Exact simulation means $\QC[A,B]=\SW[A,B]$ for every channel pair. For approximation, we use
\begin{equation}\label{eq:simulation-error}
 \operatorname{err}(\QC;d)=
 \sup_{A,B\ \mathrm{CPTP}}\frac12\norm{\QC[A,B]-\SW[A,B]}_\diamond .
\end{equation}
The diamond norm allows inputs entangled with a reference. We write $\TD(\rho,\sigma)=\frac12\norm{\rho-\sigma}_1$ for trace distance.

A simpler use of the switch prepares the control in $\ket+=(\ket0+\ket1)/\sqrt2$ and the target in $I/d$, then measures Pauli $X$ on the control. Its expectation value is the \emph{visibility}
\begin{equation}\label{eq:visibility}
 v(A,B)=\frac1d\Re\sum_{i,j}\Tr(B_jA_iB_j^\dagger A_i^\dagger).
\end{equation}
The outcomes $\pm$ have probabilities $(1\pm v)/2$. Gao et al.~\cite{gao2023} used this measurement to quantify incompatibility: $1-v=\operatorname{NCOM}_{I/d}^2$ in their notation, and
\[
 1-v(A,B)=\frac1{2d}\sum_{i,j}\norm{[A_i,B_j]}_{\rm HS}^2.
\]
Thus $v=1$ exactly when every $A_i$ commutes with every $B_j$. For projective measurements, this is equivalent to compatibility. We write $v(\mathcal E)=v(\mathcal E,\mathcal E)$ for the self-visibility used in our noise test.

\subsection{Circuits with one call to each box}\label{sec:ordered}

A circuit that calls each box at most once uses one of the two orders on each run. Both simulation proofs reduce to this case and use the following Choi representation and causal constraints.

Let $P,F$ be the circuit's input and output, and let $A_I,A_O,B_I,B_O$ be the channel-slot systems. Remove the boxes from a Kraus branch of the circuit, that is, a fixed sequence of outcomes and Kraus indices of the known instruments, to obtain a vector $v_\alpha$ on these six systems. Inserting operators $K,K'$ contracts the slot indices; we call this map the \emph{link}:
\begin{equation}\label{eq:vector-link}
 [\mathcal L(K,K')v]_{p,o}
 =\sum_{a_i,a_o,b_i,b_o}v_{p,a_i,a_o,b_i,b_o,o}\,K_{a_o,a_i}K'_{b_o,b_i}.
\end{equation}
The circuit's Choi operator is $C=\sum_\alpha v_\alpha v_\alpha^\dagger$; we call the $v_\alpha$ its Kraus vectors. Our vectorization convention is $\vket K=\sum_{a,b}K_{ba}\ket a\ket b$, with the input factor first. A channel has Choi operator $J_A=\sum_i\vket{A_i}\vbra{A_i}$ and $\Tr_{A_O}J_A=I$. Linking $C$ with $J_A\otimes J_B$ gives the output Choi operator.

\begin{lemma}[Ordered components]\label{lem:ordered}
For a circuit that calls each box at most once, pad each run to one call per box. Its Choi operator decomposes as
\[
 C=C_{A\before B}+C_{B\before A},
 \qquad C_{A\before B},C_{B\before A}\succeq0.
\]
The first term satisfies the homogeneous comb equations
\begin{equation}\label{eq:ordered-comb}
\begin{aligned}
 d\,\Tr_F C_{A\before B}&=\bigl(\Tr_{B_OF}C_{A\before B}\bigr)\otimes I_{B_O},\\
 d\,\Tr_{B_IB_OF} C_{A\before B}&=\bigl(\Tr_{A_OB_IB_OF}C_{A\before B}\bigr)\otimes I_{A_O},
\end{aligned}
\end{equation}
and the second satisfies the equations with $A$ and $B$ exchanged. For any fixed input, the two orders are chosen with probabilities $p_{A\before B}$ and $p_{B\before A}=1-p_{A\before B}$ independent of the boxes.
\end{lemma}

\begin{proof}
Group the Kraus branches by the box called first. That choice is made before either box is used, so its probabilities are independent of the channels. Each group gives a positive Choi operator for a fixed query order. The comb equations express that later outputs cannot affect earlier inputs~\cite{wechs2021}; their homogeneous form allows the group's weight to depend on the input.
\end{proof}

\section{Exact simulation: no finite number of calls suffices}\label{sec:exact}

\begin{figure}[t]
\centering
\includegraphics{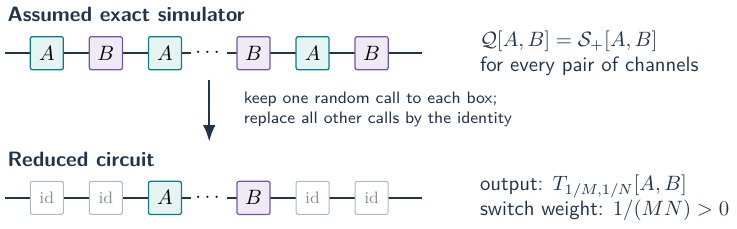}
\caption{Reducing a finite-query simulator, with $M$ calls to $A$ and $N$ calls to $B$, to one call per box. Universal correctness on mixtures with the identity forces the reduced circuit to implement a diluted switch (Lemma~\ref{lem:polarization}). Its positive switch weight gives the contradiction in Proposition~\ref{prop:no-diluted}. One possible query order is shown; known operations and memory are omitted.}
\label{fig:exact}
\end{figure}

We prove Theorem~\ref{thm:exact-simulation} by contradiction. Suppose a finite-query circuit reproduces the switch for every pair of channels, even with the control fixed to $\ket+$. Write $\SW_+[A,B]$ for this channel from the target input to the control and target output. Its Kraus operators are
\begin{equation}\label{eq:switch-plus}
 \tfrac1{\sqrt2}\bigl(\ket0\otimes B_jA_i+\ket1\otimes A_iB_j\bigr).
\end{equation}
Universal correctness includes channels mixed with the identity. We use these mixtures to turn the proposed simulator into a circuit with one call to each box. Its output contains the switch with a positive weight, which we show is impossible for qubits. Figure~\ref{fig:exact} illustrates the reduction; we extend the contradiction to every dimension at the end.

\subsection{Keeping one call to each channel}\label{sec:polarization}

For $\lambda,\mu\in[0,1]$, define the \emph{diluted switch}
\begin{equation}\label{eq:dilution}
\begin{split}
 T_{\lambda,\mu}[A,B]
 &=\SW_+\bigl[\lambda A+(1-\lambda)\id,\ \mu B+(1-\mu)\id\bigr]\\
 &=\lambda\mu\,\SW_+[A,B]
   +\lambda(1-\mu)\,\SW_+[A,\id]\\
 &\quad+(1-\lambda)\mu\,\SW_+[\id,B]
   +(1-\lambda)(1-\mu)\,\SW_+[\id,\id].
\end{split}
\end{equation}
The switch has weight $\lambda\mu$; each remaining term uses at most one nonidentity channel. We will obtain $T_{1/M,1/N}$ from an exact simulator with $M$ calls to $A$ and $N$ calls to $B$.

Number the calls to each box in the order they occur. Replacing the $k$th call to $A$ by a channel $A^{(k)}$ and the $l$th call to $B$ by $B^{(l)}$ makes the output separately linear in these $M+N$ channels. This holds even for adaptive circuits: on each fixed sequence of instrument outcomes, every numbered channel appears once in a composition of completely positive maps, and summing over outcomes preserves linearity.

For $S\subseteq\{1,\dots,M\}$ and $S'\subseteq\{1,\dots,N\}$, let $\QC_{S,S'}[A,B]$ keep only the calls indexed by $S,S'$ and replace the others by the identity. Keeping one uniformly random call to each box gives
\[
 \QC_1=\frac1{MN}\sum_{k=1}^{M}\sum_{l=1}^{N}\QC_{\{k\},\{l\}}.
\]

\begin{lemma}[Reduction to one call per box]\label{lem:polarization}
Let $M,N\ge1$ and let $\QC$ be an adaptive circuit that makes exactly $M$ calls to $A$ and $N$ calls to $B$ on every run, with $\QC[A,B]=\SW_+[A,B]$ for all channels $A,B$ on $\C^d$. Then $\QC_1[A,B]=T_{1/M,1/N}[A,B]$ for all such channels.
\end{lemma}

\begin{proof}
Let $A_x=xA+(1-x)\id$ and $B_y=yB+(1-y)\id$ with $0\le x,y\le1$. Exactness for these mixtures and separate linearity give
\[
 \SW_+[A_x,B_y]
 =\sum_{S,S'}x^{|S|}(1-x)^{M-|S|}\,
 y^{|S'|}(1-y)^{N-|S'|}\;\QC_{S,S'}[A,B].
\]
We compare coefficients of $x(1-x)^{M-1}y(1-y)^{N-1}$. These coefficients are well defined: the functions $x^a(1-x)^{M-a}y^b(1-y)^{N-b}$ are linearly independent on $(0,1)^2$, since division by $(1-x)^M(1-y)^N$ turns them into distinct monomials in $x/(1-x)$ and $y/(1-y)$.

The coefficient on the right is $MN\,\QC_1[A,B]$. On the left, expand the four terms in~\eqref{eq:dilution} and multiply by $(x+(1-x))^{M-1}(y+(1-y))^{N-1}=1$ to express them in the same basis. Their coefficients are $1$, $N-1$, $M-1$, and $(M-1)(N-1)$, respectively. For example, the $A$-only term already supplies $x(1-y)$; its factor $y$ must come from one of the $N-1$ additional factors. Thus
\[
 \begin{aligned}
 MN\,\QC_1[A,B]&=\SW_+[A,B]+(N-1)\,\SW_+[A,\id]\\
 &\quad+(M-1)\,\SW_+[\id,B]+(M-1)(N-1)\,\SW_+[\id,\id].
 \end{aligned}
\]
Dividing by $MN$ gives $T_{1/M,1/N}$.
\end{proof}

This reduction requires correctness on general channels. The known three-call simulation assumes a unitary input~\cite{chiribella2013,bavaresco2025}, and the identity mixtures above need not satisfy that promise.

\subsection{Ruling out the diluted switch}\label{sec:qubit-certificate}

The one-call impossibility result concerns the switch itself~\cite{chiribella2013,araujo2015}. Our reduction leaves it with weight $1/(MN)$, which can be arbitrarily small. We therefore need the following stronger statement.

\begin{proposition}[No diluted switch from definite orders]\label{prop:no-diluted}
Let $0<\lambda,\mu\le1$. No adaptive circuit that calls each of two qubit boxes at most once implements $T_{\lambda,\mu}[A,B]$ for all pairs of qubit channels $A,B$.
\end{proposition}

The proof has two parts. Positivity restricts the circuit to a small space of possible Kraus vectors. Within that space, a linear constraint imposed by either query order is incompatible with a positive switch weight. Suppose such a circuit exists, and pad unused calls as in Lemma~\ref{lem:ordered}.

\paragraph{Restricting the support.}
Insert unitary channels $U,V$. The four terms of the diluted switch have Kraus operators proportional to
\begin{equation}\label{eq:four-kraus}
 \ket+\otimes I,\qquad \ket+\otimes U,\qquad \ket+\otimes V,\qquad
 \tfrac1{\sqrt2}\bigl(\ket0\otimes VU+\ket1\otimes UV\bigr).
\end{equation}
Let $\mathcal H(U,V)$ be their vectorized span. If a circuit gives this output, each of its output Kraus vectors must lie in $\mathcal H(U,V)$: positive rank-one terms outside that support cannot cancel. A Kraus vector $v$ of the process has seven qubit indices, for the target input, the two slots, and the control and target output. Using the link in~\eqref{eq:vector-link}, it must therefore belong to
\[
 \mathcal V=\bigl\{v\in\C^{128}:\ \mathcal L(U,V)v\in\mathcal H(U,V)
                    \text{ for all qubit unitaries }U,V\bigr\}.
\]

\begin{lemma}[Support]\label{lem:support}
For $\lambda,\mu\in[0,1]$, every Kraus vector of a circuit with one call to each qubit box implementing $T_{\lambda,\mu}$ on all unitary pairs lies in $\mathcal V$. This space has dimension $25$ and is the range of the full-column-rank real matrix $E\in\mathbb Z^{128\times25}$ given in Appendix~\ref{sec:support-proof}.
\end{lemma}

The positivity argument above proves the first claim. The columns of $E$ describe four kinds of paths: the target bypasses both boxes, passes through $A$ only, passes through $B$ only, or undergoes the switch. They span $\mathcal V$ by the exact rank calculation in Appendix~\ref{sec:support-proof}.

\paragraph{The incompatible linear constraint.}
By Lemma~\ref{lem:ordered}, the circuit's Choi operator is $C_{A\before B}+C_{B\before A}$, with positive components satisfying the respective comb equations. The support lemma applies to the Kraus vectors in both components and gives
\[
 C_{A\before B}=EX_{AB}E^\top,\qquad
 C_{B\before A}=EX_{BA}E^\top.
\]
Here $X_{AB},X_{BA}$ are Hermitian $25\times25$ matrices.

Let $C_T$ represent the four terms of~\eqref{eq:dilution}, with each unused slot fed a maximally mixed dummy state and its output discarded. Equality of the output on all channel inputs does not require equality of the process Choi operators. For example, if a box's output is discarded, feeding it $\ket0$ or $\ket1$ changes the process Choi operator but has no effect on the final output, because the box preserves the trace. Let $\mathcal G$, the \emph{gauge}, be the space of operators that give zero output on every channel pair. Correctness means
\begin{equation}\label{eq:gauge-equation}
 C_{A\before B}+C_{B\before A}-C_T\in\mathcal G.
\end{equation}
Appendix~\ref{sec:gauge-proof} gives the linear projector that characterizes $\mathcal G$. The basis $E$, the reference operator $C_T$, and all constraint equations are real. Taking real parts therefore preserves the comb equations and~\eqref{eq:gauge-equation}, so we may take $X_{AB},X_{BA}$ to be real symmetric.

\begin{lemma}[A linear constraint on the switch weight]\label{lem:exact-obstruction}
There is a linear functional $f$ on real symmetric $25\times25$ matrices $X$ with the following properties.
\begin{enumerate}
\item[\textup{(a)}] $f(X)=0$ whenever $EXE^\top$ satisfies the $A$-before-$B$ comb equations~\eqref{eq:ordered-comb}.
\item[\textup{(b)}] $f(X)=0$ whenever $EXE^\top$ satisfies the $B$-before-$A$ comb equations.
\item[\textup{(c)}] $f(X)=0$ whenever $EXE^\top\in\mathcal G$.
\item[\textup{(d)}] For all $\lambda,\mu\in[0,1]$, we have $C_T=EX_TE^\top$ for a diagonal matrix $X_T$, and $f(X_T)=\lambda\mu/2$.
\end{enumerate}
\end{lemma}

\Needspace{4\baselineskip}
The last column of $E$ represents the switch. The functional has the form $f(X)=X_{24,24}+g(X)$, where $g$ uses only off-diagonal entries. Thus $f(X_T)$ reads the positive coefficient of the switch vector, while the two causal constraints and the gauge each force $f$ to vanish. Appendix~\ref{sec:certificate} gives the explicit functional and verifies these identities in exact arithmetic.

\begin{proof}[Proof of Proposition~\ref{prop:no-diluted}]
The two ordered terms satisfy $f(X_{AB})=f(X_{BA})=0$. By~\eqref{eq:gauge-equation} and Lemma~\ref{lem:exact-obstruction}(c,d), their difference from the diluted switch also satisfies $f(X_{AB}+X_{BA}-X_T)=0$. Thus
\[
 0=f(X_{AB})+f(X_{BA})=f(X_T)=\frac{\lambda\mu}{2}>0,
\]
a contradiction.
\end{proof}

\subsection{Completing the proof in every dimension}\label{sec:embedding}

\begin{proof}[Proof of Theorem~\ref{thm:exact-simulation}]
For qubits, pad the proposed finite-budget simulator to make $M,N\ge1$ calls to the two boxes on every run. Lemma~\ref{lem:polarization} gives a circuit with one call to each box implementing $T_{1/M,1/N}$, contradicting Proposition~\ref{prop:no-diluted}.

For $d>2$, an exact simulator would also simulate the qubit switch. Choose an isometry $J:\C^2\to\C^d$ and the decoding channel
\[
 R(\rho)=J^\dagger\rho J+
 \Tr\bigl[(I-JJ^\dagger)\rho\bigr]\,\proj0 .
\]
For a qubit channel $A$, define $A^\uparrow(\rho)=JA(R(\rho))J^\dagger$. A call to $A^\uparrow$ uses one call to $A$. Its Kraus operators are $JA_iJ^\dagger$ and $JA_i\ket0\!\bra{e_k}$, where $(e_k)$ is an orthonormal basis of the complement of the encoded subspace. All map into that subspace, and the second group vanishes on it. Thus, on an encoded input,
\[
 \SW_+[A^\uparrow,B^\uparrow]\bigl(J\rho J^\dagger\bigr)
 =(I\otimes J)\,\SW_+[A,B](\rho)\,(I\otimes J)^\dagger .
\]
Encoding the input, using the proposed simulator, and decoding the target would give an exact qubit simulator with the same query budget, which we have ruled out. Finally, a simulator for every control input must work for $\ket+$.
\end{proof}

The reduction above uses exact agreement on all identity mixtures. To bound simulation at a fixed positive error, we need the quantitative argument in the next section.

\section{Approximate simulation}\label{sec:approximation}

The hard channels mix many unknown unitaries. With few calls, a circuit rarely uses the same constituent unitary twice. We compare its output with that of a process that samples the unitaries without replacement. The one-call constraints then bound the coherence between the two orders for each pair. For every circuit $\QC$ with budget $q$, every $d\ge2$, and every integer $r\ge1$, we prove
\begin{equation}\label{eq:tradeoff-intro}
 \operatorname{err}(\QC;d)\ge
 \frac12-\frac{13r^2}{3d}-\frac{q(q-1)}{2r}.
\end{equation}
The bound already holds for pairs of uniform mixtures of $r$ unitary channels. Choosing $r$ of order $\sqrt d$ will give Theorem~\ref{thm:simulation-summary}.

\subsection{The channels and the test}\label{sec:witness}

Sample independent Haar-random unitaries $U_1,\ldots,U_r$ and $V_1,\ldots,V_r$ on $\C^d$, and define
\[
 A=\frac1r\sum_{i=1}^{r}\Ad(U_i),\qquad B=\frac1r\sum_{j=1}^{r}\Ad(V_j).
\]
Here $\Ad(U)(\rho)=U\rho U^\dagger$. Once sampled, these are fixed memoryless channels. Each call to $A$ applies a uniformly chosen $U_i$, and similarly for $B$.

Prepare the control in $\ket+$ and the target maximally entangled with a reference. For a unitary $W$, write $\ket W=d^{-1/2}\sum_k\ket k\otimes W\ket k$. The switch output on this \emph{probe} is $\rho_{\mathrm{sw}}=SS^\dagger/r^2$, where the columns of $S$ are
\[
 \ket{s_{ij}}=\tfrac1{\sqrt2}\bigl(\ket0\ket{V_jU_i}+\ket1\ket{U_iV_j}\bigr),
 \qquad 1\le i,j\le r.
\]
Let $S_0,S_1$ have columns $\ket{V_jU_i},\ket{U_iV_j}$, respectively.

We use these vectors to define a test for the output. If they were orthonormal, $SS^\dagger$ would project onto their span and accept the switch with probability one. Lemma~\ref{lem:gram} will show that their Gram matrix is close to the identity on average when $r^2\ll d$. Dividing $SS^\dagger$ by its norm when necessary gives a valid measurement effect:
\begin{equation}\label{eq:effect}
 \Theta=\frac{SS^\dagger}L,\qquad L=\max\bigl(1,\norm{SS^\dagger}_\infty\bigr),
\end{equation}
where $\norm{\cdot}_\infty$ is the operator norm. We have $0\preceq\Theta\preceq I$ and $\Theta\preceq SS^\dagger$. The test may depend on the sampled unitaries because the measurement used to distinguish two channels in the diamond norm may depend on the channels themselves.

\looseness=-1 For the circuit output $\rho$, it is convenient to upper-bound $\Tr(\Theta\rho)$ by $\Tr(SS^\dagger\rho)$. The latter separates into diagonal terms and a coherence term. In the control basis, $SS^\dagger$ has diagonal blocks $S_0S_0^\dagger/2,S_1S_1^\dagger/2$ and off-diagonal block $S_0S_1^\dagger/2$. Thus, writing the blocks of $\rho$ as $\rho_{00},\rho_{01},\rho_{11}$, we obtain
\begin{equation}\label{eq:block-form}
 \Tr(SS^\dagger\rho)=\underbrace{\tfrac12\Tr(S_0S_0^\dagger\rho_{00})
 +\tfrac12\Tr(S_1S_1^\dagger\rho_{11})}_{\text{diagonal part}}
 +\underbrace{\Re\sum_{i,j}\bra{V_jU_i}\rho_{01}\ket{U_iV_j}}_{\text{cross part}} .
\end{equation}
\looseness=-1 For the switch, both parts are close to $1/2$ on average when $r^2\ll d$. The diagonal part alone cannot separate the switch from a circuit that chooses an order classically. We therefore need to bound the cross part.

\begin{lemma}[Near-orthonormality]\label{lem:gram}
For $K=S_0$, $S_1$ and $S$,
\begin{equation}\label{eq:gram}
 \E\norm{K^\dagger K-I}_\infty\le\frac{r^2}d.
\end{equation}
Consequently, the test accepts the switch with probability
\begin{equation}\label{eq:R1}
 \E\Tr(\Theta\rho_{\mathrm{sw}})\ge1-\frac{r^2}d.
\end{equation}
\end{lemma}

\begin{proof}
The diagonal entries of $S_0^\dagger S_0$ are one. For distinct pairs $(i,j)$ and $(k,l)$, an off-diagonal entry is
\[
 \langle V_jU_i|V_lU_k\rangle=\tfrac1d\Tr(U_i^\dagger V_j^\dagger V_lU_k),
 \qquad(i,j)\ne(k,l).
\]
If $i=k$, the $U$ factors cancel under the trace and $V_j^\dagger V_l$ is Haar-random. Otherwise $U_k$ appears only once, so the product is again Haar-random. In both cases the expected squared modulus is $\E|\Tr U|^2/d^2=d^{-2}$. Summing their expected squared moduli gives $\E\norm{S_0^\dagger S_0-I}_{\mathrm F}^2=(r^4-r^2)/d^2$. The operator norm is bounded by the Frobenius norm, so Jensen's inequality proves the claim for $S_0$. The argument for $S_1$ is the same. For $S$, use $S^\dagger S=(S_0^\dagger S_0+S_1^\dagger S_1)/2$ and the triangle inequality.

For the acceptance bound, let $G=S^\dagger S$ and $\zeta=\norm{G-I}_\infty$. Then $L\le1+\zeta$. The $r^2$ eigenvalues of $G$ sum to $r^2$, so Cauchy--Schwarz gives $\Tr G^2\ge r^2$. Therefore
\[
 \Tr(\Theta\rho_{\mathrm{sw}})=\frac{\Tr G^2}{r^2L}
 \ge\frac1{1+\zeta}\ge1-\zeta .
\]
Taking expectations proves~\eqref{eq:R1}.
\end{proof}

We next bound the coherence a circuit can produce for one pair of unitaries. This will let us control the cross part of the test.

\subsection{Reversing a product with ordered calls}\label{sec:reverse}

\looseness=-1 A circuit that calls $U$ before $V$ can apply $VU$ directly. It can also produce $UV$ by gate teleportation, on a Bell-measurement outcome of probability $d^{-2}$. A bound on its ability to reverse the order must include this success probability. We therefore bound the overlap of an unnormalized output with $\ket{UV}$.

\begin{figure}[t]
\centering
\includegraphics{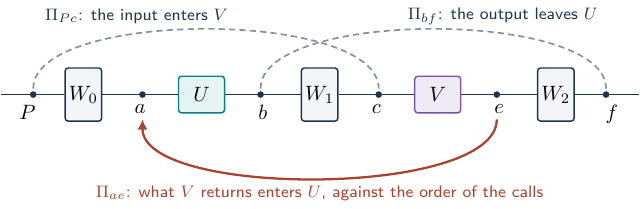}
\caption{Reversing the product when $U$ is called before $V$. The projection $\Pi_{ae}$ connects the output of the later call to the input of the earlier one. The comb equations limit its weight and give the reversed-product bound. Quantum memory between the known operations $W_t$ is omitted.}

\label{fig:wiring}
\end{figure}

Label the input, the two oracle slots, and the output by $P,(a,b),(c,e),f$, as in Figure~\ref{fig:wiring}. Each system has dimension $d$. Write $\ket\Phi=\vket I/\sqrt d$ and $\Pi=\proj\Phi$, with subscripts specifying the systems.

\begin{lemma}[Overlap with the reversed product]\label{lem:reverse}
Let $C\succeq0$ on the wires $P,a,b,c,e,f$ obey the comb equations~\eqref{eq:ordered-comb} of a circuit that calls $U$ before $V$, and let $p=\Tr C/d^3$. Let $0\preceq C'\preceq C$, and let $\sigma'(U,V)$ be the unnormalized output of $C'$ when $P$ is half of a maximally entangled state. For independent Haar-random $U,V$,
\begin{equation}\label{eq:bd}
 \E\,\bra{UV}\sigma'(U,V)\ket{UV}\ \le\ p\,b_d,
 \qquad b_d=\left(\frac1d+\frac d{d^2-1}\right)^2.
\end{equation}
\end{lemma}

For an ordered component of a circuit, $p$ is its probability on the probe, and $C'$ can select later outcomes. The overlap in~\eqref{eq:bd} therefore includes both probabilities. The bound is $O(d^{-2})$.

\begin{proof}
Write $C'=\sum_\alpha v_\alpha v_\alpha^\dagger$. Inserting $U,V$ maps each $v_\alpha$ to an operator $K:P\to f$ through the link~\eqref{eq:vector-link}. Its contribution to the overlap is $|\Tr((UV)^\dagger K)|^2/d^2$. To express this in terms of $C'$, define
\[
 \ket{z_{U,V}}=d^{-1}\,\vket{U^*}_{ab}\otimes\vket{V^*}_{ce}\otimes\vket{UV}_{Pf},
\]
where $U^*$ is the entrywise complex conjugate. The contribution above is $|\langle z_{U,V},v_\alpha\rangle|^2$; the conjugates account for the antilinearity of the inner product. Summing over $\alpha$ and averaging gives
\[
 \E\,\bra{UV}\sigma'(U,V)\ket{UV}=\Tr(C'\,\Gamma)\le\Tr(C\,\Gamma),\qquad
 \Gamma=\E\proj{z_{U,V}}\succeq0 .
\]
All dependence on the circuit is now in $C$. We first bound the operator $\Gamma$, then use the causal constraints on $C$. To average over $U$ and $V$ separately, introduce two auxiliary systems $x,x'$ at the intermediate wire of $UV$:
\[
 \vket{UV}_{Pf}=\vbra I_{xx'}\bigl(\vket V_{Px'}\otimes\vket U_{xf}\bigr),
 \qquad\text{so}\qquad
 \Gamma=d^{-2}\,\vbra I_{xx'}\bigl(\Gamma_U\otimes\Gamma_V\bigr)\vket I_{xx'},
\]
where $\Gamma_U$ is the average of the projector onto $\vket{U^*}_{ab}\otimes\vket U_{xf}$, and $\Gamma_V$ is the average of the projector onto $\vket{V^*}_{ce}\otimes\vket V_{Px'}$. Set $\ket\omega=\vket I_{ab}\otimes\vket I_{xf}$; the first vector is $(U^*\otimes U)_{bf}\ket\omega$. The representation $U^*\otimes U$ on $b,f$ fixes $\ket\Phi_{bf}$ and is irreducible on its orthogonal complement, of dimension $d^2-1$. Since $\bra\Phi_{bf}\ket\omega=\ket\Phi_{ax}$ and $\Tr_{bf}\proj\omega=I_{ax}$, Schur's lemma~\cite{collins2006} gives, with $\beta=(d^2-1)^{-1}$,
\begin{equation}\label{eq:frame}
 \Gamma_U=\Pi_{ax}\Pi_{bf}+\beta(I-\Pi_{ax})(I-\Pi_{bf})
 \preceq\Pi_{ax}\Pi_{bf}+\beta I,
\end{equation}
with the analogous bound $\Gamma_V\preceq\Pi_{Pc}\Pi_{ex'}+\beta I$. Taking tensor products and contracting with $\vket I_{xx'}$ preserves these positive-operator inequalities. Under the contraction, $\Pi_{ax}\Pi_{ex'}$ becomes $\Pi_{ae}/d$, either single projector becomes $I/d$, and the identity becomes $d$. Expanding gives
\begin{equation}\label{eq:envelope}
 \Gamma\preceq d^{-3}\bigl(\Pi_{Pc}\Pi_{ae}\Pi_{bf}
 +\beta\,\Pi_{bf}+\beta\,\Pi_{Pc}+d^2\beta^2 I\bigr).
\end{equation}
The leading term requires the reversed connection in Figure~\ref{fig:wiring}. The first comb equation gives $\Tr_f C=C_2\otimes I_e$, with $C_2=\Tr_{ef}C/d$ and $\Tr C_2=pd^2$. Using $\Tr_e\Pi_{ae}=I_a/d$, we get
\[
 \Tr(C\,\Pi_{Pc}\Pi_{ae}\Pi_{bf})
 \le\Tr(C\,\Pi_{Pc}\Pi_{ae})
 =d^{-1}\Tr(C_2\Pi_{Pc})\le pd.
\]
The other terms in~\eqref{eq:envelope} are bounded using $\Tr C=pd^3$. Thus
\[
 \Tr(C\Gamma)\le d^{-3}\bigl(pd+2\beta pd^3+d^2\beta^2pd^3\bigr)
 =p\bigl(d^{-1}+d\beta\bigr)^2=p\,b_d. \qedhere
\]
\end{proof}

Positivity turns this overlap bound into a bound on coherence: an off-diagonal matrix element is at most the geometric mean of its two diagonal entries.

\begin{corollary}[Coherence between the two orders]\label{cor:coherence}
Let $\sigma(U,V)$ be the output on the probe of an adaptive circuit that calls each of $U$ and $V$ at most once, and let $\sigma_{01}$ be its block between the control states $\ket0$ and $\ket1$. For independent Haar-random $U,V$,
\begin{equation}\label{eq:coherence}
 \left|\E\bra{VU}\sigma_{01}(U,V)\ket{UV}\right|\le\sqrt{b_d}.
\end{equation}
\end{corollary}

\begin{proof}
Lemma~\ref{lem:ordered} decomposes the output as $\sigma=\sigma^{(1)}+\sigma^{(2)}$, one term for each order, with probabilities $p_1+p_2=1$ independent of $U,V$. For the first order, positivity and Cauchy--Schwarz give
\begin{align*}
 \bigl|\E\bra{VU}\sigma^{(1)}_{01}\ket{UV}\bigr|
 &\le\sqrt{\E\bra{VU}\sigma^{(1)}_{00}\ket{VU}}
      \sqrt{\E\bra{UV}\sigma^{(1)}_{11}\ket{UV}}\\
 &\le p_1\sqrt{b_d}.
\end{align*}
The first factor is at most $\sqrt{p_1}$ by normalization. For the second, apply Lemma~\ref{lem:reverse} with $C$ obtained by discarding the control and $C'\preceq C$ by selecting control outcome $1$. The other order gives $p_2\sqrt{b_d}$ with the roles exchanged. Adding the bounds proves the claim.
\end{proof}

\subsection{Applying the bound to mixtures of unitaries}\label{sec:hidden}

Set $\eta=q(q-1)/(2r)$. If $q>r$, then $\eta\ge1$ and \eqref{eq:tradeoff-intro} is trivial, so assume $q\le r$. Compare the actual process, which draws an independent uniform index at every call, with a \emph{fresh} process that samples without replacement. For the latter, sample two independent, uniformly random sequences of $q$ distinct indices in advance and use the next entry whenever the circuit calls that box. Each constituent unitary is used at most once, although the circuit still chooses its calls adaptively. This comparison process has memory of the indices already used. Write the outputs as $\rho_{\mathrm{act}}$ and $\rho_{\mathrm{fresh}}$.

\begin{claim}[Replacing repeated indices]\label{clm:fresh}
For every choice of the unitaries, the outputs of the circuit in the two processes satisfy $\TD(\rho_{\mathrm{act}},\rho_{\mathrm{fresh}})\le\eta$.
\end{claim}

\begin{proof}
At call $t$, at most $t-1$ indices have already been used for the chosen box. The new independent index repeats one of them with probability at most $(t-1)/r$. If it does not repeat, it is uniform over the unused indices, as in the fresh process. The probability of any repetition is therefore at most $\sum_{t\le q}(t-1)/r=\eta$.

Write $\rho_{\mathrm{act}}=\tau+\tau_1$, where $\tau$ is the sum of the contributions from histories with no repeated index. Then $\Tr\tau_1\le\eta$. Every history with no repeated index has at least as much weight in the fresh process: after $m$ calls to the same box, the probability of an unused index is larger by a factor $r/(r-m)\ge1$. Hence $\tau_2=\rho_{\mathrm{fresh}}-\tau\succeq0$ and $\Tr\tau_2=\Tr\tau_1$. This gives $\TD(\rho_{\mathrm{act}},\rho_{\mathrm{fresh}}) =\frac12\norm{\tau_1-\tau_2}_1\le\eta$.
\end{proof}

\begin{claim}[Coherence for each unitary pair]\label{clm:pair}
For every pair $(i,j)$, $\bigl|\E\bra{V_jU_i}(\rho_{\mathrm{fresh}})_{01}\ket{U_iV_j}\bigr|\le\sqrt{b_d}$.
\end{claim}

\begin{proof}
Fix the two index sequences and all unitaries except $U_i,V_j$. These choices are independent of $U_i,V_j$, which remain independent and Haar-random. Every other call is now a known operation. The resulting circuit retains its adaptivity but uses each of $U_i,V_j$ at most once, so Corollary~\ref{cor:coherence} applies. Averaging over the fixed choices proves the claim.
\end{proof}

\begin{proof}[Proof of Theorem~\ref{thm:simulation-summary}]
We first bound the circuit's acceptance in the fresh process. For $k=0,1$, $\norm{S_kS_k^\dagger}_\infty =\norm{S_k^\dagger S_k}_\infty \le1+\norm{S_k^\dagger S_k-I}_\infty$. Since $\Tr\rho_{00}+\Tr\rho_{11}=1$, the diagonal part of~\eqref{eq:block-form} is at most $\frac12(1+\max_k\norm{S_k^\dagger S_k-I}_\infty)$. Bound the maximum by the sum and apply Lemma~\ref{lem:gram}; the expectation is at most $1/2+r^2/d$. For the cross part, Claim~\ref{clm:pair} contributes at most $\sqrt{b_d}$ per pair, so the total is at most $r^2\sqrt{b_d}\le7r^2/(3d)$, where we used $d\ge2$.

Since $0\preceq\Theta\preceq I$ and $\Theta\preceq SS^\dagger$, Claim~\ref{clm:fresh} gives $\Tr(\Theta\rho_{\mathrm{act}}) \le\Tr(SS^\dagger\rho_{\mathrm{fresh}})+\eta$. For each channel pair, the difference in acceptance probabilities is bounded by the trace distance of the outputs, and hence by $\operatorname{err}(\QC;d)$. Averaging and using Lemma~\ref{lem:gram} yields
\begin{align*}
 \operatorname{err}(\QC;d)
 &\ge\E\Tr(\Theta\rho_{\mathrm{sw}})-\E\Tr(\Theta\rho_{\mathrm{act}})\\
 &\ge\Bigl(1-\frac{r^2}d\Bigr)-\Bigl(\frac12+\frac{r^2}d+\frac{7r^2}{3d}\Bigr)-\eta\\
 &=\frac12-\frac{13r^2}{3d}-\frac{q(q-1)}{2r}.
\end{align*}
This proves~\eqref{eq:tradeoff-intro} using only uniform mixtures of $r$ unitaries.

For fixed $\epsilon<1/2$ and sufficiently large $d$, choose $r=\lfloor c\sqrt d\rfloor$, where $c>0$ satisfies $13c^2/3<(1/2-\epsilon)/2$. If the error is at most $\epsilon$, then~\eqref{eq:tradeoff-intro} requires $q(q-1)/(2r)>(1/2-\epsilon)/2$. Therefore $q=\Omega_\epsilon(d^{1/4})$.
\end{proof}

\subsection{Circuits for approximate simulation}\label{sec:simulation}

Measuring the control and calling the boxes in the chosen order implements $\mathcal D\circ\SW[A,B]$, where $\mathcal D$ dephases the control. This uses two calls and has error at most $\frac12\norm{\id-\mathcal D}_\diamond=1/2$.

For smaller error, use the deterministic port-based teleportation processor of Pirandola et al.~\cite{pirandola2019}. It takes $M$ copies of a channel's Choi state and approximates that channel with error at most $d(d-1)/M$, measured in half the diamond norm. Prepare the program states for both boxes with $2M$ calls, and denote the resulting channels by $\tilde A,\tilde B$. The processors have known circuits; their unitary dilations, acting on separate program registers, allow us to implement $\SW[\tilde A,\tilde B]$ by coherently controlling their order.

With either input fixed, the switch contracts diamond distance in the other input. To see this, fix $B$ and choose a unitary dilation of it. Apply this dilation before the call to $A$ when the control is $\ket1$, and after it when the control is $\ket0$, using the same environment. Discarding the environment implements $\SW[A,B]$. This is an ordinary circuit with one open slot for $A$, so processing cannot increase the diamond distance between two choices of $A$. The same argument applies with the inputs exchanged.

Replacing $A$ and then $B$ therefore gives error at most $2d(d-1)/M$. Taking $M=\lceil2d(d-1)/\epsilon\rceil$ gives a universal simulation with $O(d^2/\epsilon)$ calls for every $\epsilon>0$.

\section{Distinguishing dephasing from depolarizing noise}\label{sec:c6-proof}

We now prove Theorem~\ref{thm:c6}. The device dephases in an unknown fixed basis, while its theoretical model is depolarizing noise. For a unitary $U$, let $P_j=U\proj jU^\dagger$ and write
\begin{equation}\label{eq:conjugation}
 \Delta_U(\rho)=\sum_{j=1}^d P_j\rho P_j,
 \qquad \Delta_U=\Ad(U)\circ\Delta_I\circ\Ad(U^\dagger),
\end{equation}
and let $\Omega(\rho)=I\Tr(\rho)/d$ be complete depolarization. At noise strength $0<p\le1$, the two hypotheses are
\begin{equation}\label{eq:validation-pair}
 \mathcal E_{p,U}=(1-p)\id+p\,\Delta_U,
 \qquad
 \mathcal M_p=(1-p)\id+p\,\Lambda,
 \qquad
 \Lambda=\E_U\Delta_U=\frac{\id+d\,\Omega}{d+1}.
\end{equation}
The actual basis $U$ stays fixed on every call. In the model, the Haar average is taken separately for each use. The second moment of a Haar-random unit vector $\psi$ gives $\Lambda(\rho)=d\,\E_\psi\bra\psi\rho\ket\psi\,\proj\psi =(\rho+I\Tr\rho)/(d+1)$.

Both channels are unital and, for every fixed $U$, have the same entanglement fidelity $1-p+p/d$ and average gate fidelity $1-p(d-1)/(d+1)$ relative to the identity. Under gate-independent, time-independent Markovian noise, randomized benchmarking estimates the latter quantity~\cite{magesan2011}. The device can therefore agree with its model on these benchmarks while acting differently on individual states.

Dephasing preserves its basis states, whereas the depolarizing model mixes them. If the basis were known, the circuit could prepare one of these states and test for this mixing. Here the basis is unknown. Since $\E_U\mathcal E_{p,U}=\mathcal M_p$, averaging over Haar-random $U$ gives identical outputs for every one-call experiment. Repeated calls can reveal correlations because the basis stays fixed. Our lower bound allows the circuit to use these correlations, retain quantum outputs, and adapt later inputs. The switch instead tests whether the Kraus operators commute, without learning the basis.

\subsection{A test using the quantum switch}\label{sec:switch-test}

\begin{figure}[ht]
\centering
\includegraphics{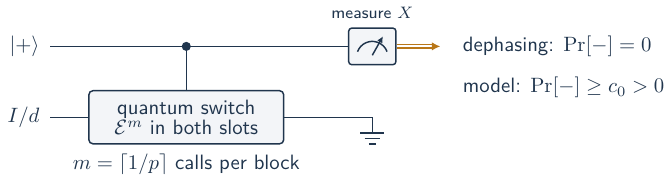}
\caption{Testing the noise model with the switch. Repeated calls within each block make the model produce $-$ with probability at least an absolute constant $c_0$, while dephasing never does. A constant number of repetitions uses $O(1/p)$ device calls.}
\label{fig:geometry}
\end{figure}

We use the control measurement of Gao et al.~\cite{gao2023}, with both slots of the switch filled by the device, as in Figure~\ref{fig:geometry}. With control $\ket+$ and target $I/d$, the expectation of $X$ on the control is
\begin{equation}\label{eq:validation-visibility}
 v(\mathcal E_{p,U})=1,
 \qquad
 v(\mathcal M_p)=1-p^2\,\frac{d-1}{d+1}.
\end{equation}
The first equality follows because the Kraus operators $\sqrt{1-p}\,I$ and $\sqrt p\,P_j$ commute. For the second, write $\mathcal M_p=\alpha\id+(1-\alpha)\Omega$, where $\alpha=1-pd/(d+1)$. The visibility from Section~\ref{sec:visibility} is linear in each channel, equals $1$ if either channel is the identity, and satisfies $v(\Omega,\Omega)=1/d^2$ by substituting the Kraus operators $\ket x\!\bra y/\sqrt d$. Thus $v(\mathcal M_p)=1-(1-\alpha)^2(1-d^{-2})$.

The outcome $-$ never occurs for dephasing noise, and occurs with probability $p^2(d-1)/(2(d+1))$ for the model. To avoid a cost of $O(1/p^2)$ when the noise is weak, we let it accumulate: each slot now contains a block of $m=\lceil1/p\rceil$ consecutive device calls. For dephasing, $\mathcal E_{p,U}^m$ is still a mixture of the identity and $\Delta_U$, so its Kraus operators still commute. For the model, the block channel and the probability of $-$ are
\[
 \mathcal M_p^m=\alpha^m\id+(1-\alpha^m)\Omega,
 \qquad
 \Pr[-]=\frac12(1-\alpha^m)^2(1-d^{-2}).
\]
For $d\ge2$, $\alpha^m\le e^{-2/3}$, so this probability is bounded below by an absolute positive constant. We repeat the experiment a constant number of times and accept the model if any outcome is $-$. This succeeds with probability at least $2/3$ for the model and with certainty for every fixed dephasing basis, using $O(1/p)$ device calls. This proves the upper bound in Theorem~\ref{thm:c6}.

An adaptive circuit can also run this test using the approximate simulation from Section~\ref{sec:simulation}. Constant amplification and a sufficiently small simulation error give success probability at least $2/3$ using $O(d^2)$ calls to the block channel, hence $O(d^2/p)$ device calls.

\subsection{Reducing the problem to two measurements}\label{sec:pair}

We turn to the lower bound and take $d=2^n$ throughout. We first consider $p=1$, where the two channels are $\Delta_U$ and $\Lambda$, and extend the result to other strengths in the next subsection. It is easier to work with the completely depolarizing channel $\Omega$ in place of $\Lambda$. This changes the channel by only $\frac12\norm{\Lambda-\Omega}_\diamond\le1/(d+1)$, since $\Lambda-\Omega=(\id-\Omega)/(d+1)$.

The reason for this replacement is that $\Delta_U$ and $\Omega$ can both be written as two successive measurements:
\[
 \Delta_U=\Delta_U\circ\Delta_U,\qquad
 \Omega=\Delta_{UH}\circ\Delta_U,\qquad H=H_2^{\otimes n},
\]
where $H_2$ is the $2\times2$ Hadamard matrix and $H$ is the Walsh--Hadamard transform, so the bases defined by $U$ and $UH$ are mutually unbiased. We now give the circuit separate access to the two measurements, allowing it to choose which box to call next. The two possibilities are
\begin{equation}\label{eq:c6-priors}
 (A,B)=(\Delta_U,\Delta_U)\quad\text{(same-basis case)}
 \qquad\text{or}\qquad
 (A,B)=(\Delta_U,\Delta_{UH})\quad\text{(unbiased case)},
\end{equation}
where $U$ is drawn from the Haar measure once and kept fixed. Any circuit making $L$ calls to $\Delta_U$ or $\Omega$ can be run with these boxes using $2L$ calls: replace each device call by a call to $A$ followed by one to $B$. Thus a lower bound for the two-box problem gives a lower bound for the original device.

Fix an adaptive circuit with budget $q$ for the two boxes, and let $\rho_0$ and $\rho_1$ be its final states on the output and memory in the same-basis and unbiased cases, averaged over $U$. A circuit that succeeds for every fixed basis must also succeed on this average, so it suffices to bound the distance between these states.

\begin{lemma}[Distinguishability of the two measurement pairs]\label{lem:c6-comparison}
There is an absolute constant $C$ such that, for $q\ge1$ and $4q\le d$,
\begin{equation}\label{eq:c6-comparison}
 \TD(\rho_0,\rho_1)\le C\,\frac{q^2}{d^{1/8}}.
\end{equation}
\end{lemma}

The proof records the queries to $U$ and $U^\dagger$ and simplifies this record to a process we call the label cache. The recording error is $O(q^2/d^{1/8})$, and the simplification adds $O(q^2/\sqrt d)$. The two resulting caches differ by only $O(q^2/d)$. The first error therefore determines the exponent in our lower bound. Before proving the lemma, we use it to handle arbitrary noise strength.

\subsection{Reduction from weak noise to full-strength noise}\label{sec:validation}

Noise of strength $p$ can be simulated by applying the corresponding channel of strength one with probability $p$ and doing nothing otherwise. A circuit with $q$ weak-noise calls then uses at most $pq$ full-strength calls on average. We turn this into a fixed query budget by stopping the simulation if it makes too many calls.

\begin{proof}[Proof of the lower bound in Theorem~\ref{thm:c6}]
Let $L=\lfloor a d^{1/16}\rfloor$, where $a>0$ is an absolute constant chosen below. We first show that a circuit with at most $L$ calls cannot distinguish $\Delta_U$ from $\Lambda$ with success probability greater than $7/12$, when the alternatives are equally likely and $U$ is Haar-random.

If $L=0$, the success probability is at most $1/2$. Otherwise, replace $\Lambda$ by $\Omega$ and implement each device call by the two measurement boxes. Choose $a\le1/8$, so $8L\le d$ and Lemma~\ref{lem:c6-comparison} applies with budget $2L$. Replacing $\Omega$ back by $\Lambda$ adds at most $L/(d+1)$ to the trace distance. Thus the final states for $\Delta_U$ and $\Lambda$ obey
\[
 \TD(\rho_{\Delta},\rho_{\Lambda})
 \le C\,\frac{(2L)^2}{d^{1/8}}+\frac{L}{d+1}
 \le 4Ca^2+a.
\]
Choose $a$ small enough that this is at most $1/6$. The optimal success probability is then at most $\frac12(1+1/6)=7/12$.

Now suppose a circuit with budget $q$ distinguishes $\mathcal M_p$ from every $\mathcal E_{p,U}$ with success probability at least $2/3$. Simulate its calls using a full-strength device, applying that device with probability $p$ at each call. This reproduces both channels exactly and uses at most $pq$ device calls in expectation. Stop before the $(L+1)$st full-strength call and output any guess. Markov's inequality bounds the stopping probability by $pq/(L+1)$ in either case. The stopped circuit has budget $L$, so the bound above gives
\[
 \frac23-\frac{pq}{L+1}\le\frac7{12},
 \qquad\text{hence}\qquad
 q\ge\frac{L+1}{12p}>\frac{a}{12}\,\frac{d^{1/16}}p. \qedhere
\]
\end{proof}

We now prove Lemma~\ref{lem:c6-comparison}.

\subsection{The label cache}\label{sec:cache}

\begin{figure}[!b]
\centering
\includegraphics{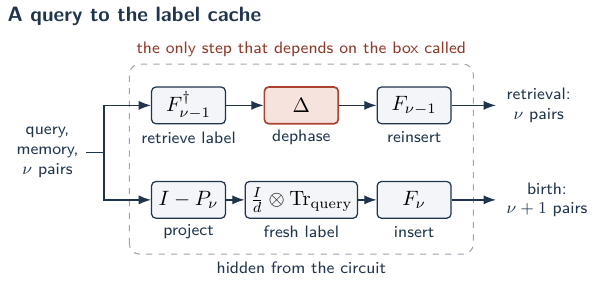}
\caption{A call to the label cache. The cache either retrieves a stored label, with effect $P_\nu$, or creates a fresh label, with effect $I-P_\nu$. Only retrieval depends on the measurement basis. Replacing $\Delta$ by $\Omega$ gives the reference cache used in Section~\ref{sec:invariant}.}
\label{fig:cache}
\end{figure}

\looseness=-1 A circuit can keep the output of a measurement and return it in a later call. To account for what it learns this way, write each box as $U^\dagger$, followed by dephasing in a known basis, followed by $U$, as in~\eqref{eq:conjugation}. The known basis is computational, except for box $B$ in the unbiased case, where it is the Hadamard basis. The choice of this middle measurement is the only difference between the two cases.

The recording theorem of Ma and Huang~\cite{mahuang2025} approximates queries to a Haar-random unitary by a process that stores pairs of quantum registers. One register, the \emph{label}, holds the input to a forward query; the other, the \emph{image}, records its relation to the returned output. Both are hidden from the circuit. The theorem keeps separate lists for forward and inverse queries. Proposition~\ref{prop:cache} reduces them to a single list of symmetric pairs, which we call the label cache.

A cache call either retrieves a stored label, dephases it, and reinserts it, or creates a fresh maximally mixed label. We call these actions \emph{retrieval} and \emph{birth}. Births do not depend on the measurement basis, since every dephasing channel preserves $I/d$. We therefore only need to compare retrievals. The labels and images remain quantum registers, so the cache retains their correlations with earlier outputs.

\begin{definition}[Label cache]\label{def:cache}
A \emph{pair} consists of a $d$-dimensional label register and a $d$-dimensional image register. For $\nu\ge0$, let $\mathcal H_\nu=\operatorname{Sym}^\nu(\C^d\otimes\C^d)$ be the symmetric space of $\nu$ pairs, and let $a^\dagger_{xy}$ be the bosonic creation operator from $\mathcal H_\nu$ to $\mathcal H_{\nu+1}$ for a pair with label $\ket x$ and image $\ket y$, normalized so that $[a_{xy},a^\dagger_{x'y'}]=\delta_{xx'}\delta_{yy'}$. For $0\le\nu<d$, define
\[
 E_\nu=\frac1{\sqrt d}\sum_{x,y}\ket y\bra x\otimes a^\dagger_{xy},
 \qquad
 F_\nu=E_\nu\bigl(E_\nu^\dagger E_\nu\bigr)^{-1/2},
\]
acting from the query system and $\mathcal H_\nu$ to the query system and $\mathcal H_{\nu+1}$. Define the projectors $P_\nu=F_{\nu-1}F_{\nu-1}^\dagger$ for $\nu\ge1$, with $P_0=0$.

The \emph{label cache} maintains a classical counter $\nu$, initially $0$, and a state in $\mathcal H_\nu$, both hidden from the circuit. For a dephasing channel $\Delta=\Delta_V$ in a known basis, a call at counter value $\nu$ applies the following two-outcome instrument to the query and the stored pairs:
\begin{equation}\label{eq:c6-cache}
\begin{aligned}
 \mathcal R_{\Delta,\nu}(X)
 &=F_{\nu-1}(\Delta\otimes\id)
       (F_{\nu-1}^\dagger XF_{\nu-1})F_{\nu-1}^\dagger,\\
 \mathcal B_\nu(X)
 &=F_\nu\left[\frac Id\otimes\Tr_{\rm query}
       ((I-P_\nu)X(I-P_\nu))\right]F_\nu^\dagger.
\end{aligned}
\end{equation}
The retrieval map $\mathcal R_{\Delta,\nu}$ leaves the counter unchanged and is zero when $\nu=0$. The birth map $\mathcal B_\nu$ increments the counter to $\nu+1$.
\end{definition}

The map $E_\nu$ stores the incoming query as a new label and makes the outgoing query maximally entangled with the new image, then symmetrizes the pairs. The normalization in $F_\nu$ makes it an isometry: Lemma~\ref{appA:gram} gives
\[
 E_\nu^\dagger E_\nu
 =I+\frac1d\sum_{i\le\nu}
 \operatorname{SWAP}(\text{query},\text{label}_i)
 \succeq(1-\nu/d)I,
\]
which is invertible for $\nu<d$. The adjoint $F_{\nu-1}^\dagger$ retrieves a label from the range of $P_\nu$, while a birth acts on the complementary subspace; see Figure~\ref{fig:cache}.

On the first call, $P_0=0$, so the cache creates a new label. Since $F_0\ket x=d^{-1/2}\sum_y\ket y\otimes\ket{x,y}$, the outgoing query is maximally entangled with the stored image. If the circuit returns that query unchanged, the next call retrieves the stored label with certainty. Thus the cache accounts for reuse of earlier outputs, even though the circuit never sees the stored pairs.

The traces of the two maps are $\Tr\mathcal R_{\Delta,\nu}(X)=\Tr(P_\nu X)$ and $\Tr\mathcal B_\nu(X)=\Tr((I-P_\nu)X)$, so each cache call preserves trace. It acts only on the query and the stored pairs, so the circuit's memory is untouched. The counter is at most $i-1$ before call $i$.

\paragraph{Rotating the labels.}
A property of $F_\nu$ will let us compare the two caches: rotating the incoming label and all stored labels before insertion has the same effect as rotating all stored labels afterward. For every unitary $T$ on $\C^d$,
\begin{equation}\label{eq:covariance}
 \bigl(I\otimes T^{\otimes(\nu+1)}\bigr)F_\nu
 =F_\nu\bigl(T\otimes T^{\otimes\nu}\bigr),
\end{equation}
where $T^{\otimes\nu}$ acts on the stored labels, and the additional $T$ on the right acts on the incoming query. The images and the outgoing query are unchanged. To verify this identity, first substitute the definition of $E_\nu$. It follows that $E_\nu^\dagger E_\nu$ commutes with $T\otimes T^{\otimes\nu}$, as does its inverse square root, so the identity also holds for $F_\nu$. We will also use the consequence that $P_\nu$ commutes with $I\otimes T^{\otimes\nu}$.

\begin{proposition}[Approximating the boxes by a label cache]\label{prop:cache}
Let $q\ge1$ and $4q\le d$, and fix an adaptive circuit with budget $q$. In either case of~\eqref{eq:c6-priors}, let $\rho$ be its final state, averaged over $U$, and let $\rho^{\mathrm{cache}}$ be its final state when the label cache answers every call. The cache uses $\Delta_I$ for box $A$, and $\Delta_I$ or $\Delta_H$ for box $B$ in the same-basis or unbiased case, respectively. Then
\begin{equation}\label{eq:eps-cache}
 \TD(\rho,\rho^{\mathrm{cache}})
 \le\varepsilon_{\mathrm{rec}}+\varepsilon_{\mathrm{cache}},
 \qquad
 \varepsilon_{\mathrm{rec}}=O\!\left(\frac{q^2}{d^{1/8}}\right),
 \qquad
 \varepsilon_{\mathrm{cache}}=O\!\left(\frac{q^2}{\sqrt d}\right),
\end{equation}
with absolute constants in both $O(\cdot)$ terms.
\end{proposition}

The first term is the error in the recording theorem of Ma and Huang. The second comes from converting their two lists into the cache. Appendix~\ref{sec:c6-recording} proves this conversion, including the correlations with the circuit's memory.

\subsection{Comparing the two caches}\label{sec:invariant}

The two caches differ only in the basis used to dephase a retrieved label. We compare each with a \emph{reference cache} that uses $\Omega$ in place of $\Delta$ and is the same for both cases. Its joint state is invariant under a common rotation of all labels. For labels with this symmetry, dephasing one of them has almost the same effect as depolarizing it. The next two lemmas establish these facts. We then replace the calls one at a time, always comparing them on a state produced by the reference cache. This avoids assuming a symmetry that the original caches need not have.

\begin{lemma}[The reference cache is invariant]\label{lem:reference-invariant}
Let an adaptive circuit with budget at most $d$ interact with the reference cache. At every step, conditional on each counter value $\nu$, the joint state of the query, memory, and stored pairs is invariant under a common rotation $T^{\otimes\nu}$ of the stored labels, for every unitary $T$.
\end{lemma}

\begin{proof}
We prove the claim by induction over the operations of the experiment. There are no stored labels initially, and operations of the circuit act only on its query and memory, so they preserve the claimed symmetry.

Suppose the state before a cache call is $X$, with counter value $\nu$. A retrieval can occur only for $\nu\ge1$, in which case covariance~\eqref{eq:covariance} implies that $X'=F_{\nu-1}^\dagger XF_{\nu-1}$ is invariant under $T\otimes T^{\otimes(\nu-1)}$. The first factor rotates the query, which now holds the retrieved label. Depolarizing this label preserves the symmetry, and inserting it again with $F_{\nu-1}$ returns a state invariant under $T^{\otimes\nu}$. For a birth, $P_\nu$ commutes with the rotation and the fresh label $I/d$ is invariant. Applying $F_\nu$ then gives invariance under $T^{\otimes(\nu+1)}$, as required.
\end{proof}

\looseness=-1 With one stored label, this symmetry forces the label to be maximally mixed and independent of the rest of the system. Dephasing and depolarizing it therefore give the same state. Several labels can still be correlated. The next lemma bounds how much dephasing and depolarization can then differ.

\begin{lemma}[Dephasing invariant labels]\label{lem:invariant-label}
Let $X\succeq0$ be an operator of trace $s$ on $k\ge1$ labels of dimension $d\ge2$ and a reference system $Z$. If $X$ is invariant under $T^{\otimes k}\otimes I_Z$ for every unitary $T$, then every dephasing channel $\Delta=\Delta_V$ on the first label satisfies
\begin{equation}\label{eq:invariant-label}
 \norm{(\Delta\otimes\id)(X)-(\Omega\otimes\id)(X)}_1
 \le\frac{6(k-1)}d\,s.
\end{equation}
\end{lemma}

\begin{proof}
We may take $s=1$ by homogeneity and use the computational basis by unitary invariance. The first label is maximally mixed, so a measurement of it gives each value $j$ with probability $1/d$. Let $\sigma_j=d\bra jX\ket j$ be the conditional state of the remaining labels and $Z$ after outcome $j$, and write $\bar\sigma=d^{-1}\sum_j\sigma_j$. Then
\[
 (\Delta\otimes\id)(X)=\frac1d\sum_j\proj j\otimes\sigma_j,
 \qquad
 (\Omega\otimes\id)(X)=\frac Id\otimes\bar\sigma .
\]
The difference is block diagonal, with trace norm $d^{-1}\sum_j\norm{\sigma_j-\bar\sigma}_1$. We therefore compare the conditional states $\sigma_j$ and $\sigma_a$ for two distinct outcomes $j$ and $a$.

Let $P_{ja}$ project onto the subspace where none of the remaining labels equals $j$ or $a$. Exchanging these two basis values acts as the identity on this subspace. Since $X$ is invariant under that exchange, $P_{ja}\sigma_jP_{ja}=P_{ja}\sigma_aP_{ja}$. Invariance under diagonal unitaries also implies that $\sigma_j$ commutes with the operators counting each basis value, and hence with $P_{ja}$; the same holds for $\sigma_a$. Their difference is therefore supported on the complementary subspace, giving
\[
 \norm{\sigma_j-\sigma_a}_1
 \le\Tr((I-P_{ja})\sigma_j)+\Tr((I-P_{ja})\sigma_a).
\]

To bound these two traces, fix one of the remaining labels. Its joint state with the first label is invariant under $T\otimes T$, and the projector onto equal basis values twirls to $(I+\operatorname{SWAP})/(d+1)$. The two labels thus coincide with probability at most $2/(d+1)$. By symmetry under basis permutations, this is also an upper bound on the conditional probability that the other label equals $j$ when the first is $j$. The other $d-1$ values are equally likely, so each specified value $a\ne j$ has conditional probability at most $1/(d-1)$. A union bound over the $k-1$ remaining labels gives
\[
 \Tr((I-P_{ja})\sigma_j)
 \le(k-1)\left(\frac2{d+1}+\frac1{d-1}\right),
\]
and the same bound holds for $\sigma_a$. Averaging the bound on $\norm{\sigma_j-\sigma_a}_1$ over $a$, with zero contribution from $a=j$, gives
\[
 \norm{\sigma_j-\bar\sigma}_1
 \le\frac{2(k-1)(3d-1)}{d(d+1)}\le\frac{6(k-1)}d. \qedhere
\]
\end{proof}

\begin{proof}[Proof of Lemma~\ref{lem:c6-comparison}]
Fix one of the two cases. We replace its cache calls by reference cache calls one at a time, starting with the first. For $0\le i\le q$, let $\tau_i$ be the circuit's final state when the first $i$ calls use the reference cache and the rest use the cache for this case. Thus $\tau_0=\rho^{\mathrm{cache}}$ from Proposition~\ref{prop:cache}, and $\tau_q$ is the final state for the reference cache, common to both cases.

The experiments giving $\tau_{i-1}$ and $\tau_i$ differ only at call $i$, and then only if a label is retrieved. Fix the classical history before that call and the counter value $\nu\le i-1$, and let $X$ be the corresponding unnormalized state. There is no retrieval when $\nu=0$. Otherwise, the first $i-1$ calls used the reference cache, so Lemma~\ref{lem:reference-invariant} and~\eqref{eq:covariance} show that $X'=F_{\nu-1}^\dagger XF_{\nu-1}$ is invariant under a common rotation of its $\nu$ labels. Conditioning on the history preserves this invariance because its projector acts only on the circuit memory. The $\nu$ labels are the retrieved label, now in the query, and the $\nu-1$ labels still in the cache.

We can now apply Lemma~\ref{lem:invariant-label} with $k=\nu$, taking the stored images and circuit memory as the reference $Z$. This is valid because $\mathcal H_{\nu-1}$ is a subspace of $\nu-1$ labels and $\nu-1$ images. Changing $\Omega$ to the dephasing channel used on this history changes the state by at most $6(\nu-1)\Tr X'/d\le6(q-1)\Tr X'/d$ in trace norm. The weights $\Tr X'\le\Tr X$ sum to at most $1$ over all histories and counter values, and later operations cannot increase the trace distance. Hence $\TD(\tau_{i-1},\tau_i)\le3(q-1)/d$. Summing over the calls gives
\[
 \TD\bigl(\rho^{\mathrm{cache}},\tau_q\bigr)\le\frac{3q(q-1)}d
 \qquad\text{for each of the two cases.}
\]
Both caches are therefore close to the same reference state. Combining this bound with Proposition~\ref{prop:cache} and the triangle inequality gives
\[
 \TD(\rho_0,\rho_1)
 \le2\bigl(\varepsilon_{\mathrm{rec}}+\varepsilon_{\mathrm{cache}}\bigr)
 +\frac{6q(q-1)}d .
\]
Since $\varepsilon_{\mathrm{rec}}=O(q^2/d^{1/8})$ and $\varepsilon_{\mathrm{cache}}=O(q^2/\sqrt d)$, the right-hand side is at most $Cq^2/d^{1/8}$ for an absolute constant $C$. This proves \eqref{eq:c6-comparison}.
\end{proof}

\section{Discussion}\label{sec:discussion}\label{sec:where}\label{sec:open}

The noise-model decision gives a query advantage for the switch in a task motivated by quantum computing. Dephasing in an unknown basis and its depolarizing model have the same entanglement and average gate fidelities, yet the switch distinguishes them through the incompatibility measurement of Gao et al.~\cite{gao2023}. This suggests looking for similar advantages in more structured noise families. Our hard instances allow an arbitrary unknown basis; identifying physically motivated families that retain the separation would bring the result closer to noise characterization in practice.

\looseness=-1 The information returned by a device matters. If projective measurements report their outcomes, an ordinary circuit can estimate $v(A,B)$ by applying $A$, then $B$, then $A$ to $I/d$ and comparing the two outcomes of $A$~\cite{gao2023}. Those outcomes can therefore make the scalar accessible without the switch. General simulation is a different demand: our approximate lower bound still holds when each random-unitary call reports its selected index, provided the circuit cannot choose that index or obtain the unitary's description. The proofs of Claims~\ref{clm:fresh} and~\ref{clm:pair} already allow such classical records.

There is also room to sharpen the query bounds. For the noise-model decision, the exponent $1/16$ comes from the error in recording forward and inverse Haar-unitary queries~\cite{mahuang2025}; a direct analysis of measurement channels may give a stronger bound. For simulation, it would be useful to understand how the cost depends on noise strength and Kraus rank, between the three-call construction for a unitary input~\cite{bavaresco2025} and the mixtures of about $\sqrt d$ unitaries used here.

\paragraph{Concurrent and independent work.}\label{sec:concurrent}
\looseness=-1 The three main results were obtained in September 2026 and are present in the repository's initial October 2026 commit~\cite{repo}. The repository at \mbox{\url{https://github.com/dtlics/Indefinite-Causal-Order}} contains the proof code, exact certificates, and accompanying Lean formalization, built on Lean-QIT~\cite{leanqit2026}. While preparing this manuscript, we learned of independent work by Cheng~\cite{cheng2026} and Goldin~\cite{goldin2026}. Cheng also rules out finite-query exact simulation and extends the result to switches of more than two channels. Both exact proofs use channel mixtures to reduce to one call per input; ours then uses the linear certificate in Lemma~\ref{lem:exact-obstruction}. Goldin proves an $\Omega(d^{1/3})$ lower bound for approximate simulation, improving the exponent of our bound. His bound comes from a one-bit discrimination problem that one use of the switch decides. It uses nonunital channels that send a distinguished state to a uniform mixture of hidden orthogonal states and send every state orthogonal to the distinguished state back to it. Our noise-model decision concerns unital dephasing and depolarizing channels, so that lower bound does not establish Theorem~\ref{thm:c6}.

\paragraph{Use of AI.}
The author chose the research directions and central theme of this paper, drawing on prior literature. The theory and writing were developed jointly with AI systems from the GPT, Claude, and Gemini families. The author guided the work and suggested high-level techniques, while the AI systems worked out most of the technical details. AI also developed the Lean formalization.

\appendix
\section{Reduction to the label cache}\label{sec:c6-recording}

\subsection*{Setup}\label{appA:setting}

We prove Proposition~\ref{prop:cache} by converting the recording process of Ma and Huang~\cite{mahuang2025} into the label cache. We retain the hidden records and the circuit's memory when comparing intermediate processes. Their correlations are therefore included in the bounds.

We use Theorem~8 of~\cite{mahuang2025}, equation~(9.5), for $d=2^n$; the operators $V^L,V^R,V,E^L,E^R$ from that paper; their partial-isometry properties in Claims~14 and~15; and the estimate in Claim~22. We prove the remaining steps below.

\paragraph{The purified experiment.}
Fix either case in~\eqref{eq:c6-priors} and an adaptive circuit with budget $q$. Dilate its known instruments to Stinespring isometries $W_i$ on the query and memory, with $W_0$ preparing the input. They record the classical outcomes and may depend on earlier ones. Keep the discarded environments to purify the experiment, without allowing later operations to act on them. If the circuit stops early, set its output aside and pad the remaining calls using box $A$ on a dummy input. Every run then has exactly $q$ calls. Let $\pi^c_i$ project onto classical records that select box $c$ at call $i$.

Use~\eqref{eq:conjugation} to write each box as $U^\dagger$, a known measurement, and $U$. The known basis $\psi^c_1,\dots,\psi^c_d$ is computational for box $A$, and computational or Hadamard for box $B$ in the two cases. Retaining the box choice and measurement outcome in a fresh hidden register $O_i$ gives the isometry
\[
 \widehat\Delta_i=\sum_{c\in\{A,B\}}\sum_{j=1}^d
   \pi^c_i\otimes\proj{\psi^c_j}\otimes\ket{c,j}_{O_i}.
\]
For fixed $U$, the purified experiment and its averaged output are
\begin{equation}\label{appA:experiment}
 \ket{\Psi_U}=W_q\,\bigl(U\widehat\Delta_qU^\dagger\bigr)\,W_{q-1}\cdots
   \bigl(U\widehat\Delta_1U^\dagger\bigr)\,W_0\ket0,
 \qquad
 \rho=\E_U\Tr_O\proj{\Psi_U},
\end{equation}
where $O=O_1\cdots O_q$. We retain the query and all memory after $W_q$, since discarding unused registers can only reduce trace distance. Define $\rho^{\mathrm{cache}}$ the same way with the cache in place of the boxes. For vectors that also contain the lists introduced below, abbreviate $\TD(\Psi,\Psi')=\TD(\Tr_{OLR}\proj\Psi,\Tr_{OLR}\proj{\Psi'})$.

\paragraph{Registers and notation.}
The record has two hidden lists of label--image pairs: $L$ for forward queries and $R$ for inverse queries. Each query operator can insert or delete a pair. A list of $\nu$ pairs lies in $\operatorname{Sym}^\nu(\C^d\otimes\C^d)$, with an orthonormal basis indexed by multisets of pairs.

A \emph{history} $\eta=(c_1,j_1,\dots,c_i,j_i)$ lists the boxes and hidden outcomes from the first $i$ calls. It corresponds to the basis vector $\ket\eta$ of $O_1\cdots O_i$. Its \emph{outcome vectors} are $\psi_s=\psi^{c_s}_{j_s}$; write $\mathcal K_s=\operatorname{span}\{\psi_1,\dots,\psi_s\}$, with $\mathcal K_0=\{0\}$. Let $\Pi_{\mathcal K}$ be the projector onto a subspace $\mathcal K\subseteq\C^d$. Bars denote complex conjugation in the computational basis. We need $\overline{\mathcal K}$ and $\overline\Pi_{\mathcal K}$ because inverse queries store conjugate labels; see~\eqref{appA:action}. Let $\Pi_{\le t}$ project onto lists with at most $t$ pairs in total, and set $\Pi_{\le t}=0$ for $t<0$.

Throughout, $d=2^n$, $q\ge1$ and $4q\le d$. The $2q$ unitary queries create at most $2q<d$ pairs, keeping all denominators positive and Gram operators invertible.

\begin{lemma}[Hybrid argument]\label{appA:hybrid}
Let $M_1,\dots,M_m$ and $M'_1,\dots,M'_m$ be contractions, let $v_0$ be a vector of norm at most one, and put $v_i=M'_iv_{i-1}$. Then
\[
 \norm{M_m\cdots M_1v_0-M'_m\cdots M'_1v_0}\le
 \sum_{i=1}^m\norm{(M_i-M'_i)\,v_{i-1}} .
\]
Moreover, for vectors $u,v$ of norm at most one and every partial trace $\Tr'$, $\TD(\Tr'\proj u,\Tr'\proj v)\le\norm{u-v}$.
\end{lemma}

\begin{proof}
Telescope the difference as $\sum_iM_m\cdots M_{i+1}(M_i-M'_i)v_{i-1}$ and use contractivity. The second bound follows from $\proj u-\proj v=\ket u\bra{u-v}+\ket{u-v}\bra v$, whose trace norm is at most $2\norm{u-v}$, and contractivity under partial trace.
\end{proof}

\paragraph{The five steps.}
Step~1 replaces the unknown unitary by a record of its queries. Step~2 removes the deletion terms from forward queries, whose effect is small after dephasing. Step~3 replaces the insertion operators by isometries that respect label rotations, which we need to compare the two measurement bases. Step~4 restricts births to directions orthogonal to the earlier outcome vectors. In step~5, this restriction lets us trace out the right list and identify the remaining process with the label cache. Write $\Psi_k$ for the purified experiment after the operator replacement in step~$k$, before tracing out any records. Both lists start empty, and the replacements leave the known isometries $W_i$ unchanged. Thus call~$i$ of $\Psi_1,\dots,\Psi_5$ is $V\widehat\Delta_iV^\dagger$, $V^L\widehat\Delta_iV^\dagger$, $F^L\widehat\Delta_i\widetilde V^\dagger$, $F^L\widehat\Delta_i\widehat J_i$ and $F^L\widehat\Delta_i\widehat J^{\mathrm{raw}}_i$, respectively, with the operators defined below.

\subsection{Recording the unitary and simplifying forward queries}\label{sec:c6-A1}

\paragraph{Step 1: replace the Haar unitary by a recording process.}
The recording process of~\cite{mahuang2025} acts on the query and the lists $L,R$. When the lists contain fewer than $d$ pairs in total, its insertion operators are
\begin{equation}\label{eq:c6-source-creators}
\begin{aligned}
 V^L\ket{x}\ket{L}\ket{R}
 &=\sum_{y\notin\Img(L\cup R)}
    \frac{\ket{y}\ket{L\cup\{(x,y)\}}\ket{R}}
    {\sqrt{d-|\Img(L\cup R)|}},\\
 V^R\ket{y}\ket{L}\ket{R}
 &=\sum_{x\notin\Dom(L\cup R)}
    \frac{\ket{x}\ket{L}\ket{R\cup\{(x,y)\}}}
    {\sqrt{d-|\Dom(L\cup R)|}},
\end{aligned}
\end{equation}
where $\Dom$ and $\Img$ denote the sets of labels and images, respectively. Both operators vanish on longer lists. The complete recording operator is
\begin{equation}\label{eq:c6-recording-V}
 V=V^L\bigl(I-V^RV^{R\dagger}\bigr)
   +\bigl(I-V^LV^{L\dagger}\bigr)V^{R\dagger}.
\end{equation}

Theorem~8 of~\cite{mahuang2025} allows $t$ queries to $U$ and $U^\dagger$ in a fixed pattern, with arbitrary unitaries between them. Replacing $U,U^\dagger$ by $V,V^\dagger$ and tracing out the lists changes the Haar-averaged final state by at most $9t(t+1)d^{-1/8}$ in trace distance.

Our experiment~\eqref{appA:experiment} has this form with $t=2q$: each box call uses $U^\dagger$ and then $U$. The operations $W_0,\widehat\Delta_1,W_1,\dots$ between queries are independent of $U$. They are isometries, and become unitaries if their fresh registers are included in the workspace in fixed initial states. The final $W_q$ preserves trace distance, and tracing out $O$ cannot increase it. Thus
\[
 \TD\bigl(\rho,\Tr_{OLR}\proj{\Psi_1}\bigr)\le
 \frac{9(2q)(2q+1)}{d^{1/8}}=\varepsilon_{\mathrm{rec}} .
\]

\paragraph{Step 2: remove deletion terms from forward queries.}
We want each forward query to add a pair, so we replace $V$ by $V^L$. To bound the difference, use the dephasing immediately before the query. In each term with hidden record $(c,j)$, the query is in the fixed state $\psi_j^c$ and factors from the lists. The next lemma bounds the overlap of such a state with the range of $V^R$, which controls both terms of $V-V^L$.

\begin{lemma}[Overlap with a fixed query state]\label{appA:compression}
For every unit vector $\psi\in\C^d$ and every $t<d$,
\begin{equation}\label{eq:c6-delete-compression}
 \norm{(\bra\psi\otimes I)\,\Pi_{\le t}V^RV^{R\dagger}\Pi_{\le t}\,
       (\ket\psi\otimes I)}_\infty
 \le\frac{t}{d-t+1}.
\end{equation}
\end{lemma}

\begin{proof}
Since $V^R$ adds one pair, $\Pi_{\le t}V^RV^{R\dagger}\Pi_{\le t}=V^R\Pi_{\le t-1}V^{R\dagger}$. Define $C_\psi=(\bra\psi\otimes I)V^R\Pi_{\le t-1}$. The operator in~\eqref{eq:c6-delete-compression} is $C_\psi C_\psi^\dagger$, so its norm is $\norm{C_\psi}^2$.

Let $z=\sum z(y,L,R)\ket y\ket L\ket R$ be supported on lists with at most $t-1$ pairs. Writing $R'$ for the right list after insertion, equation~\eqref{eq:c6-source-creators} gives
\[
 C_\psi z=\sum_{L,R'}\ket L\ket{R'}
 \sum_{(x,y)}\frac{\overline{\psi_x}\,z\bigl(y,L,R'\setminus\{(x,y)\}\bigr)}
 {\sqrt{d-|\Dom(L\cup R'\setminus\{(x,y)\})|}} ,
\]
where the inner sum includes the pairs $(x,y)\in R'$ for which $x$ appears neither in $L$ nor in another pair of $R'$. There are at most $t$ such pairs, and every denominator is at least $\sqrt{d-t+1}$. Applying Cauchy--Schwarz to the inner sum yields
\[
 \norm{C_\psi z}^2\le\frac{t}{d-t+1}
 \sum_{L,R'}\sum_{(x,y)}|\psi_x|^2\,
 \bigl|z\bigl(y,L,R'\setminus\{(x,y)\}\bigr)\bigr|^2
 \le\frac{t}{d-t+1}\norm z^2 .
\]
For the last inequality, regroup by the argument $(y,L,R)$ of $z$. Each argument occurs once for each $x\notin\Dom(L\cup R)$, and the sum of the corresponding $|\psi_x|^2$ is at most one.
\end{proof}

\begin{lemma}[Replacing a forward query]\label{appA:forward}
Let $\xi=\ket\psi\otimes\xi'$, where $\psi$ is a unit vector of the query and $\xi'$ is a vector of the lists and the memory with at most $t<d$ pairs. Then $\norm{(V-V^L)\xi}^2\le\frac{2t}{d-t+1}\norm\xi^2$.
\end{lemma}

\begin{proof}
By~\eqref{eq:c6-recording-V}, $V-V^L=-V^LV^RV^{R\dagger}+(I-V^LV^{L\dagger})V^{R\dagger}$. The two terms have orthogonal ranges, so
\begin{align*}
 \norm{(V-V^L)\xi}^2
 &=\norm{V^LV^RV^{R\dagger}\xi}^2
   +\norm{(I-V^LV^{L\dagger})V^{R\dagger}\xi}^2\\
 &\le2\,\bra\xi V^RV^{R\dagger}\ket\xi
 \le\frac{2t}{d-t+1}\norm\xi^2.
\end{align*}
The first inequality uses contractivity and the fact that $V^RV^{R\dagger}$ is a projector; the second is Lemma~\ref{appA:compression}.
\end{proof}

\paragraph{Error from step 2.}
Only the $q$ forward queries differ between $\Psi_1$ and $\Psi_2$. Just before forward query $i$, the state in $\Psi_2$ is $\sum_{c,j}\ket{c,j}_{O_i}\otimes\ket{\psi^c_j}\otimes\xi'_{c,j}$. There have been at most $2i-1$ queries, so each $\xi'_{c,j}$ has at most $2i-1$ pairs. Lemma~\ref{appA:forward} bounds the error on each of these orthogonal terms. Since all operators are contractions, Lemma~\ref{appA:hybrid} sums the errors to give
\begin{equation}\label{eq:c6-E}
 \TD(\Psi_1,\Psi_2)
 \le\sum_{i=1}^q\sqrt{\frac{2(2i-1)}{d-2i+2}}=\varepsilon_2.
\end{equation}

\subsection{Step 3: exactly covariant isometries}\label{sec:c6-A2}

We need the cache to respect changes of basis exactly; this is the covariance used in Section~\ref{sec:invariant}. The operator $V^L$ preserves norms on the states we use, but is only approximately covariant. The operator $E^L$ from~\cite{mahuang2025} has exact covariance but does not preserve norms. Taking its polar isometry gives both properties. We use it to replace $V^L$ by $F^L$ and $V^\dagger$ by $\widetilde V^\dagger$, as defined below.

\paragraph{Creation operators.}
Let $a^\dagger_{xy}$ and $b^\dagger_{xy}$ create a pair $(x,y)$ in the left and right lists, respectively, as in Definition~\ref{def:cache}. For a label vector $u\in\C^d$, define $a^\dagger(u,y)=\sum_xu_xa^\dagger_{xy}$ and $b^\dagger(u,y)=\sum_xu_xb^\dagger_{xy}$. These operators create a pair with label $u$ and image $y$. Following Definition~37 of~\cite{mahuang2025}, set
\begin{equation}\label{eq:c6-raw-creators}
 E^L=\frac1{\sqrt d}\sum_{x,y}\ket y\bra x\otimes a_{xy}^\dagger,
 \qquad
 E^R=\frac1{\sqrt d}\sum_{x,y}\ket x\bra y\otimes b_{xy}^\dagger .
\end{equation}
On $\nu$ left pairs, $E^L$ is $E_\nu$ from Definition~\ref{def:cache}. By~\eqref{eq:c6-raw-creators}, for $u\in\C^d$ and a list state $\Phi$,
\begin{equation}\label{appA:action}
 E^L\bigl(\ket u\otimes\Phi\bigr)
 =\frac1{\sqrt d}\sum_y\ket y\otimes a^\dagger(u,y)\Phi,
 \qquad
 (\bra u\otimes I)E^R
 =\frac1{\sqrt d}\sum_y\bra y\otimes b^\dagger(\overline u,y).
\end{equation}
The right-list label in the second identity is $\overline u$. The next lemma bounds the Gram operators, which lets us normalize $E^L$ and $E^R$ without changing them much. Its projected identity~\eqref{appA:projected} will also control births when we exclude the span of earlier outcome vectors in steps~4 and~5.

\begin{lemma}[Gram operators]\label{appA:gram}
On lists with $\nu$ left and $r$ right pairs,
\begin{equation}\label{eq:c6-grams}
 E^{L\dagger}E^L=I+\frac1d\sum_{a=1}^{\nu}
                    \operatorname{SWAP}^{\mathrm{lab}}_a,
 \qquad
 E^{R\dagger}E^R=I+\frac1d\sum_{a=1}^{r}
                    \operatorname{SWAP}^{\mathrm{im}}_a ,
\end{equation}
where $\operatorname{SWAP}^{\mathrm{lab}}_a$ exchanges the query with the label of the $a$th left pair and $\operatorname{SWAP}^{\mathrm{im}}_a$ exchanges it with the image of the $a$th right pair. More generally, for every projector $\Pi$ on $\C^d$,
\begin{equation}\label{appA:projected}
 E^{R\dagger}(\Pi\otimes I)E^R=\frac{\Tr\Pi}{d}\,I
 +\frac1d\sum_{a=1}^{r}\overline\Pi_{(a)}\,
   \operatorname{SWAP}^{\mathrm{im}}_a ,
\end{equation}
where $\overline\Pi_{(a)}$ acts on the label of the $a$th right pair. In particular, both Gram operators lie between $(1-t/d)I$ and $(1+t/d)I$ on lists with at most $t$ pairs.
\end{lemma}

\begin{proof}
By~\eqref{eq:c6-raw-creators}, the left side of \eqref{appA:projected} is $\frac1d\sum\Pi_{x'x}\ket{y'}\bra y\otimes b_{x'y'}b^\dagger_{xy}$. Insert the commutation relation $b_{x'y'}b^\dagger_{xy}=\delta_{xx'}\delta_{yy'}+b^\dagger_{xy}b_{x'y'}$. The first term gives $(\Tr\Pi/d)I$. The second sums over the right pairs. On each pair, it applies $\overline\Pi$ to the label and $\ket y\bra{y'}$ to the image. Summing this together with $\ket{y'}\bra y$ on the query exchanges the query with that image. This proves~\eqref{appA:projected}.

Set $\Pi=I$ to obtain the second identity in~\eqref{eq:c6-grams}. The same calculation with labels in place of images gives the first. Each sum contains at most $t$ operators of norm one, giving the bounds.
\end{proof}

\paragraph{Polar isometries.}
By Lemma~\ref{appA:gram}, both Gram operators are invertible when the lists have fewer than $d$ pairs. On these lists, define
\[
 F^L=E^L\bigl(E^{L\dagger}E^L\bigr)^{-1/2},\qquad
 F^R=E^R\bigl(E^{R\dagger}E^R\bigr)^{-1/2},\qquad
 P^L=F^LF^{L\dagger},\qquad P^R=F^RF^{R\dagger},
\]
and set $F^L$ and $F^R$ to zero on longer lists. They are partial isometries, and preserve norms on all states used below because those states have at most $2q<d$ pairs. On $\nu$ left pairs, $F^L=F_\nu$ and $P^L=P_\nu$ as in Definition~\ref{def:cache}.

Since $F^L-E^L=F^L\bigl(I-(E^{L\dagger}E^L)^{1/2}\bigr)$ and $|1-\sqrt\lambda|\le|1-\lambda|$, Lemma~\ref{appA:gram} gives $\norm{(F^L-E^L)\Pi_{\le t}}_\infty\le t/d$, and likewise for $F^R$. Claim~22 of~\cite{mahuang2025}, their equation~(10.25), states that $\norm{(V^L-E^L)\Pi_{\le t}}_\infty$ and $\norm{(V^R-E^R)\Pi_{\le t}}_\infty$ are at most $\sqrt{2t(t+1)/d}$. Hence
\begin{equation}\label{eq:c6-delta}
 \norm{(V^L-F^L)\Pi_{\le t}}_\infty,\
 \norm{(V^R-F^R)\Pi_{\le t}}_\infty\ \le\
 \delta_t=\sqrt{\frac{2t(t+1)}d}+\frac td .
\end{equation}
Replace $V^L,V^R$ by $F^L,F^R$ in the formula for $V^\dagger$: the new inverse query is
\begin{equation}\label{eq:c6-polar-W}
 \widetilde V^\dagger=\bigl(I-P^R\bigr)F^{L\dagger}+F^R\bigl(I-P^L\bigr).
\end{equation}
The two terms act on orthogonal input spaces and have orthogonal outputs. Each is a contraction, so their sum is also a contraction.

\begin{lemma}[Error from polar normalization]\label{appA:polar}
For every $t<d$, $\norm{(V^\dagger-\widetilde V^\dagger)\Pi_{\le t}}_\infty\le6\,\delta_t$.
\end{lemma}

\begin{proof}
Subtract~\eqref{eq:c6-polar-W} from the adjoint of \eqref{eq:c6-recording-V} and group the difference into four terms:
\begin{align*}
 V^\dagger-\widetilde V^\dagger
 ={}&\bigl(I-V^RV^{R\dagger}\bigr)\bigl(V^L-F^L\bigr)^\dagger
   +\bigl(P^R-V^RV^{R\dagger}\bigr)F^{L\dagger}\\
   &+V^R\bigl(P^L-V^LV^{L\dagger}\bigr)
   +\bigl(V^R-F^R\bigr)\bigl(I-P^L\bigr).
\end{align*}
All insertion operators increase the pair count by one, so $M^\dagger\Pi_{\le t}=(M\Pi_{\le t-1})^\dagger$ for each such operator $M$ and for their differences. Their range projectors preserve the count. Thus~\eqref{eq:c6-delta} also bounds the adjoint differences on lists with at most $t$ pairs by $\delta_t$. For the range projectors, use
\[
 AA^\dagger-BB^\dagger
 =(A-B)A^\dagger+B(A-B)^\dagger.
\]
The adjoints reduce the pair count, so the same estimate bounds this difference by $2\delta_t$ for $(A,B)=(V^L,F^L)$ or $(V^R,F^R)$. The four terms above therefore have norms at most $\delta_t$, $2\delta_t$, $2\delta_t$, and $\delta_t$, respectively. Their sum is at most $6\delta_t$.
\end{proof}

\paragraph{Error from step 3.}
The states in $\Psi_3$ contain at most $2q$ pairs. At each call, Lemma~\ref{appA:polar} bounds the change in the inverse query by $6\delta_{2q}$, and~\eqref{eq:c6-delta} bounds the change in the forward query by $\delta_{2q}$. All operators are contractions, so Lemma~\ref{appA:hybrid} gives $\TD(\Psi_2,\Psi_3)\le7q\,\delta_{2q}=\varepsilon_3$.

\subsection{Step 4: restricting births to fresh directions}\label{sec:c6-A3}

The cache treats retrieval and birth as separate outcomes. In $\widetilde V^\dagger$, however, they are two coherent terms that can interfere at a later call. To separate them, we require each new label to be orthogonal to the span $\mathcal K$ of the earlier outcome vectors. We therefore replace the inverse query by
\begin{equation}\label{eq:c6-fresh-J}
 J_{\mathcal K}=F^{L\dagger}
   +\bigl(I-\Pi_{\mathcal K}\bigr)F^R\bigl(I-P^L\bigr),
 \qquad
 \widetilde V^\dagger-J_{\mathcal K}
 =-P^RF^{L\dagger}+\Pi_{\mathcal K}F^R\bigl(I-P^L\bigr),
\end{equation}
where $\Pi_{\mathcal K}$ acts on the query. In $J_{\mathcal K}$, $F^{L\dagger}$ reverses an insertion on the range of $P^L$ and retrieves a label. The other term, a birth, acts on the complementary subspace, stores the query as a right-list image, and produces a label in $\mathcal K^\perp$. After we replace $F^R$ by $E^R$ in step~5, the right list will record which calls were births for each fixed outcome history. Claim~\ref{appA:record} shows that tracing it out then removes interference between different birth and retrieval branches.

\paragraph{The inverse query for each history.}
Let $\Theta_{\mathcal K}$ project onto lists with all left labels in $\mathcal K$ and all right labels in $\overline{\mathcal K}$. On $\nu$ left pairs and $r$ right pairs, it applies $\Pi_{\mathcal K}^{\otimes\nu}$ to the left labels and $\overline\Pi_{\mathcal K}^{\otimes r}$ to the right labels. For call $i$ in $\Psi_4$, choose the inverse query according to the history $\eta$ of the first $i-1$ calls:
\begin{equation}\label{appA:fresh-step}
 \widehat J_i=\sum_{\eta}\proj\eta_{O_1\cdots O_{i-1}}\otimes
   J_{\mathcal K_{i-1}(\eta)}\,\Theta_{\mathcal K_{i-1}(\eta)} .
\end{equation}
The projector $\Theta$ has two useful properties: it makes $\widehat J_i$ a contraction on all states, and it leaves the states in our experiment unchanged. We prove these facts in Lemma~\ref{appA:follow} and Claim~\ref{appA:support}.

\begin{lemma}[Preservation of label subspaces]\label{appA:follow}
For every projector $\Pi$ on $\C^d$ and every $\nu<d$, on lists with $\nu$ left pairs $F^L\bigl(\Pi\otimes\Pi^{\otimes\nu}\bigr)=\Pi^{\otimes(\nu+1)}F^L$, where the tensor powers act on the left labels. Consequently, for every subspace $\mathcal K$,
\begin{equation}\label{appA:follow-eq}
 F^L\bigl(\Pi_{\mathcal K}\otimes\Theta_{\mathcal K}\bigr)
   =\Theta_{\mathcal K}F^L,\qquad
 \bigl(\Pi_{\mathcal K}\otimes\Theta_{\mathcal K}\bigr)F^{L\dagger}
   =F^{L\dagger}\Theta_{\mathcal K},\qquad
 P^L\Theta_{\mathcal K}=\Theta_{\mathcal K}P^L ,
\end{equation}
and $J_{\mathcal K}\Theta_{\mathcal K}$ is a contraction.
\end{lemma}

\begin{proof}
By the first identity of~\eqref{appA:action} and $\Pi^{\otimes(\nu+1)}a^\dagger(u,y)=a^\dagger(\Pi u,y)\Pi^{\otimes\nu}$, the operator $E^L$ satisfies $E^L(\Pi\otimes\Pi^{\otimes\nu})=\Pi^{\otimes(\nu+1)}E^L$. With the adjoint identity this gives $E^{L\dagger}E^L(\Pi\otimes\Pi^{\otimes\nu}) =E^{L\dagger}\Pi^{\otimes(\nu+1)}E^L =(\Pi\otimes\Pi^{\otimes\nu})E^{L\dagger}E^L$. Its inverse square root also commutes with $\Pi\otimes\Pi^{\otimes\nu}$, so the same identity holds for $F^L$. This is~\eqref{eq:covariance} with a projector in place of a unitary. Since $F^L$ does not act on the right list, we obtain the first identity in~\eqref{appA:follow-eq}. The second is its adjoint, and the third follows by multiplication.

For the contraction bound, let $\phi=\Theta_{\mathcal K}\phi$. By the second identity in~\eqref{appA:follow-eq}, the query of the retrieval output lies in $\mathcal K$, while the query of the birth output lies in $\mathcal K^\perp$, so the two terms are orthogonal. The retrieval term has norm $\norm{P^L\phi}$, and the birth term has norm at most $\norm{(I-P^L)\phi}$. Therefore, $\norm{J_{\mathcal K}\phi}^2\le\norm{P^L\phi}^2+\norm{(I-P^L)\phi}^2 =\norm\phi^2$.
\end{proof}

\paragraph{Retrieval and birth branches.}
Fix a history $\eta$ of length $i$. A string of \emph{flags} $\beta\in\{0,1\}^i$ specifies which term of $J_{\mathcal K}$ we use at each call: $\beta_s=1$ means birth and $\beta_s=0$ means retrieval. Set $\phi_0=\ket0\ket\varnothing_L\ket\varnothing_R$ and, for $s=1,\dots,i$, define
\begin{equation}\label{appA:branches}
 \phi_s=F^L\bigl(\proj{\psi_s}\otimes I\bigr)\,J^{\beta_s}_s\,
   \Theta_{\mathcal K_{s-1}}\,\pi^{c_s}_sW_{s-1}\,\phi_{s-1},
 \qquad
 J^0_s=F^{L\dagger},\quad
 J^1_s=\bigl(I-\Pi_{\mathcal K_{s-1}}\bigr)F^R\bigl(I-P^L\bigr).
\end{equation}
After $i$ calls, the component of $\Psi_4$ with history $\eta$ is $\sum_\beta\phi_i(\eta,\beta)$. This follows from the definition of $\widehat\Delta_s$ and~\eqref{appA:fresh-step}. The same formulas with $E^R$ in place of $F^R$ describe the branches of $\Psi_5$.

\begin{claim}[Support of the stored labels]\label{appA:support}
For every history $\eta$ of length $i$ and every $\beta\in\{0,1\}^i$, $\Theta_{\mathcal K_i}\phi_i(\eta,\beta)=\phi_i(\eta,\beta)$. The same holds with $E^R$ in place of $F^R$. In particular, the projectors $\Theta$ in~\eqref{appA:fresh-step} and~\eqref{appA:branches} act as the identity.
\end{claim}

\begin{proof}
We induct on $i$. The case $i=0$ is immediate because the lists are empty. Suppose $\Theta_{\mathcal K}\phi_{i-1}=\phi_{i-1}$, where $\mathcal K=\mathcal K_{i-1}$, and set $\mathcal K'=\mathcal K_i$. Then $\mathcal K\subseteq\mathcal K'$ and $\psi_i\in\mathcal K'$. Since $W_{i-1}$ and $\pi^{c_i}_i$ do not act on the lists, $\phi=\pi^{c_i}_iW_{i-1}\phi_{i-1}$ is also fixed by $\Theta_{\mathcal K}$.

For a retrieval, $\beta_i=0$. The second identity in \eqref{appA:follow-eq} shows that $F^{L\dagger}\phi$ is fixed by $I\otimes\Theta_{\mathcal K}$ and hence by $I\otimes\Theta_{\mathcal K'}$.

For a birth, $\beta_i=1$. The third identity gives $\Theta_{\mathcal K}(I-P^L)\phi=(I-P^L)\phi$. By~\eqref{eq:c6-grams}, the factor $(E^{R\dagger}E^R)^{-1/2}$ in $F^R$ acts only on the query and right images; it leaves the labels in the same subspaces. Put $u=(I-\Pi_{\mathcal K})\psi_i$. By~\eqref{appA:action}, $(\bra{\psi_i}(I-\Pi_{\mathcal K})\otimes I)E^R$ adds a right pair with label $\overline u\in\overline{\mathcal K'}$ and leaves the other labels unchanged. The output is fixed by $\Theta_{\mathcal K'}$.

In both cases, the final $F^L$ inserts $\psi_i\in\mathcal K'$. The first identity in~\eqref{appA:follow-eq} completes the induction.
\end{proof}

We now bound the two changes in~\eqref{eq:c6-fresh-J}: removing $I-P^R$ from the retrieval term and projecting the birth label onto $\mathcal K^\perp$.

\begin{lemma}[Error from restricting new labels]\label{appA:fresh-cost}
Let $\mathcal K$ have dimension $\kappa$ and let $r<d$. On lists with $r$ right pairs,
\begin{equation}\label{eq:c6-right-compression}
 \norm{(\Pi_{\mathcal K}\otimes I)F^R}_\infty
 \le\sqrt{\frac{\kappa+r}{d-r}} .
\end{equation}
Consequently, for every $\phi=\Theta_{\mathcal K}\phi$ with at most $r$ right pairs,
\[
 \norm{(\widetilde V^\dagger-J_{\mathcal K})\phi}\le
 \left(\sqrt{\frac{\kappa+r-1}{d-r+1}}
       +\sqrt{\frac{\kappa+r}{d-r}}\right)\norm\phi ,
\]
where the first term is absent for $r=0$.
\end{lemma}

\begin{proof}
By~\eqref{appA:projected} with $\Pi=\Pi_{\mathcal K}$, $E^{R\dagger}(\Pi_{\mathcal K}\otimes I)E^R\le\frac{\kappa+r}{d}I$, since each of the $r$ terms of the sum has norm at most one. By Lemma~\ref{appA:gram}, $(E^{R\dagger}E^R)^{-1}\le(1-r/d)^{-1}I$. Conjugating the first bound with $(E^{R\dagger}E^R)^{-1/2}$ gives~\eqref{eq:c6-right-compression}.

For the second bound, the difference between the birth terms is $\Pi_{\mathcal K}F^R(I-P^L)\phi$. Equation \eqref{eq:c6-right-compression} bounds its norm by $\sqrt{(\kappa+r)/(d-r)}\,\norm\phi$. The difference between the retrieval terms is
\[
 P^RF^{L\dagger}\phi
 =F^RF^{R\dagger}(\Pi_{\mathcal K}\otimes I)F^{L\dagger}\phi ,
\]
where Lemma~\ref{appA:follow} lets us insert the projector. On $r$ right pairs, $F^{R\dagger}(\Pi_{\mathcal K}\otimes I)$ is the adjoint of $(\Pi_{\mathcal K}\otimes I)F^R$ on $r-1$ right pairs. It is zero for $r=0$; otherwise its norm is at most $\sqrt{(\kappa+r-1)/(d-r+1)}$.

Each operator changes the number of right pairs by a fixed amount, so inputs with different pair counts have orthogonal outputs. The norm on lists with at most $r$ right pairs is therefore the largest of the norms for fixed counts. Both bounds increase with that count, so the same bounds hold for at most $r$ pairs.
\end{proof}

\paragraph{Error from step 4.}
Only the inverse queries differ between $\Psi_3$ and $\Psi_4$. By Claim~\ref{appA:support}, each history component of $\Psi_4$ before inverse query $i$ is fixed by $\Theta_{\mathcal K_{i-1}}$. Here $\mathcal K_{i-1}$ has dimension $\kappa\le i-1$, and there have been at most $r\le i-1$ births. Thus we can apply Lemma~\ref{appA:fresh-cost}. Its bound increases with both parameters, so setting $\kappa=r=i-1$ bounds the error at call $i$. The first call has no error. All operators are contractions, including $\widehat J_i$ by Lemma~\ref{appA:follow}, so Lemma~\ref{appA:hybrid} gives
\begin{equation}\label{eq:c6-B}
 \TD(\Psi_3,\Psi_4)\le
 \sum_{i=2}^q\left(
       \sqrt{\frac{2i-3}{d-i+2}}
       +\sqrt{\frac{2(i-1)}{d-i+1}}\right)=\varepsilon_4.
\end{equation}

\subsection{Step 5: tracing out the right list}\label{sec:c6-A4}

The factor $(E^{R\dagger}E^R)^{-1/2}$ in $F^R$ can change images already in the right list. We remove it by replacing $F^R$ with $E^R$. Existing right pairs then remain untouched, and tracing out the list leaves the cache with outcome-dependent birth weights. We finish by replacing these weights by one.

\paragraph{Part 1: replace polar births by raw births.}
Define $J^{\mathrm{raw}}_{\mathcal K}$ and $\widehat J^{\mathrm{raw}}_i$ by replacing $F^R$ with $E^R$ in \eqref{eq:c6-fresh-J} and~\eqref{appA:fresh-step}.

\begin{lemma}[Raw births]\label{appA:raw}
Let $\mathcal K$ have dimension $\kappa$. On the right lists whose labels all lie in $\overline{\mathcal K}$,
\begin{equation}\label{eq:c6-raw-fresh-gram}
 E^{R\dagger}\bigl((I-\Pi_{\mathcal K})\otimes I\bigr)E^R
 =\Bigl(1-\frac{\kappa}{d}\Bigr)I .
\end{equation}
Consequently $J^{\mathrm{raw}}_{\mathcal K}\Theta_{\mathcal K}$ is a contraction, and $\norm{(J_{\mathcal K}-J^{\mathrm{raw}}_{\mathcal K})\phi}\le (r/d)\norm\phi$ for every $\phi$ with at most $r$ right pairs and fewer than $d$ pairs in total.
\end{lemma}

\begin{proof}
Set $\Pi=I-\Pi_{\mathcal K}$ in~\eqref{appA:projected}. Its trace is $d-\kappa$. In every other term, $I-\overline\Pi_{\mathcal K}$ acts on a right label and gives zero. This proves~\eqref{eq:c6-raw-fresh-gram}.

As in Lemma~\ref{appA:follow}, the retrieval and birth outputs are orthogonal. By~\eqref{appA:follow-eq}, $(I-P^L)\Theta_{\mathcal K}\phi$ is fixed by $\Theta_{\mathcal K}$, so~\eqref{eq:c6-raw-fresh-gram} shows that birth does not increase its norm. This proves the contraction bound. Finally, $J_{\mathcal K}-J^{\mathrm{raw}}_{\mathcal K} =(I-\Pi_{\mathcal K})(F^R-E^R)(I-P^L)$. On lists with $r'$ right pairs the spectrum of $E^{R\dagger}E^R$ lies in $[1-r'/d,\,1+r'/d]$ by~\eqref{eq:c6-grams}, whatever the left list, so $\norm{F^R-E^R}_\infty\le r'/d\le r/d$.
\end{proof}

Lemmas~\ref{appA:follow} and~\ref{appA:raw} show that the operators in $\Psi_4$ and $\Psi_5$ are contractions. Before call $i$, the right list contains at most $i-1$ pairs. Lemma~\ref{appA:hybrid} therefore gives $\TD(\Psi_4,\Psi_5)\le\sum_{i=1}^q(i-1)/d=q(q-1)/(2d)$.

\paragraph{Part 2: identify the birth record.}
\looseness=-1 From now on, $\phi_s(\eta,\beta)$ is a branch of $\Psi_5$: use~\eqref{appA:branches} with $E^R$ in place of $F^R$. Each birth adds a pair in a new label direction. This permanently records which calls were births, removing interference between different birth patterns when the right list is traced out.

Fix a history $\eta$ of length $q$ and set $u_s=(I-\Pi_{\mathcal K_{s-1}})\psi_s$. Whenever $u_s\ne0$, define $w_s=\norm{u_s}^2$ and $v_s=\overline{u_s}/\norm{u_s}$. Each $u_s$ is perpendicular to all earlier outcome vectors, so the vectors $v_s$ are orthonormal. Let $N_s=\sum_yb^\dagger(v_s,y)\,b(v_s,y)$ count the right pairs with label $v_s$, where $b(v,y)$ is the adjoint of $b^\dagger(v,y)$. Equation~\eqref{appA:action} gives
\begin{equation}\label{appA:birth-amplitude}
 \bigl(\bra{\psi_s}(I-\Pi_{\mathcal K_{s-1}})\otimes I\bigr)E^R
 =\sqrt{\frac{w_s}{d}}\;\sum_y\bra y\otimes b^\dagger(v_s,y).
\end{equation}
If $u_s=0$, the left-hand side is zero.

\begin{claim}[Permanent record of births]\label{appA:record}
Let $i\le q$ and $\beta\in\{0,1\}^i$, and let $\phi_i=\phi_i(\eta,\beta)$. \begin{enumerate}\renewcommand{\labelenumi}{(\roman{enumi})}\setlength\itemsep{0pt}
\item If $\beta_s=1$ and $u_s=0$ for some $s\le i$, then $\phi_i=0$.
\item Otherwise $N_s\phi_i=\beta_s\phi_i$ for every $s\le i$ with $u_s\ne0$, and $N_s\phi_i=0$ for every $s>i$ with $u_s\ne0$.
\end{enumerate}
Thus tracing out the right list removes all cross terms between branches with different flags:
\begin{equation}\label{appA:no-cross}
 \Tr_R\,\Bigl|\sum_\beta\phi_i(\eta,\beta)\Bigr\rangle
 \Bigl\langle\sum_\beta\phi_i(\eta,\beta)\Bigr|
 =\sum_\beta\Tr_R\proj{\phi_i(\eta,\beta)} .
\end{equation}
\end{claim}

\begin{proof}
We induct on $i$, starting with an empty right list. At call $i$, $W_{i-1}$, $\pi^{c_i}_i$, $F^L$, $F^{L\dagger}$, $P^L$ and $\proj{\psi_i}$ do not act on this list. We can omit $\Theta$ by Claim~\ref{appA:support}. A retrieval therefore commutes with every $N_s$ and leaves the number of pairs in each direction unchanged.

For a birth, equation~\eqref{appA:birth-amplitude} makes $(\proj{\psi_i}\otimes I)J^1_i$ zero when $u_i=0$. Otherwise, after $I-P^L$, it acts as $\sqrt{w_i/d}\,\ket{\psi_i}\sum_y\bra y\otimes b^\dagger(v_i,y)$. Since the $v_s$ are orthonormal, $b^\dagger(v_i,y)$ commutes with $N_s$ for $s\ne i$ and increases $N_i$ by one. Since $N_i\phi_{i-1}=0$ by induction, we obtain $N_i\phi_i=\phi_i$.

For distinct flags $\beta,\beta'$, either one branch vanishes or they have different eigenvalues under some $N_s$. Since $N_s$ is Hermitian and acts only on $R$, the branches occupy orthogonal right-list subspaces. Their cross term vanishes under $\Tr_R$, proving~\eqref{appA:no-cross}.
\end{proof}

\paragraph{The reduced weighted process.}
We now compute what remains on the query, left list and memory. At call $s$, the weight is $w_s=\norm{(I-\Pi_{\mathcal K_{s-1}})\psi_s}^2\in[0,1]$. For a positive operator $X$ on these registers, set $X'=\pi^{c_s}_sX\pi^{c_s}_s$ and define
\begin{equation}\label{appA:weighted}
\begin{aligned}
 \mathcal C^{w}_{s,0}(X)&=F^L\bigl(\proj{\psi_s}\otimes I\bigr)F^{L\dagger}
   \,X'\,F^L\bigl(\proj{\psi_s}\otimes I\bigr)F^{L\dagger},\\
 \mathcal C^{w}_{s,1}(X)&=\frac{w_s}{d}\,
   F^L\Bigl[\proj{\psi_s}\otimes
   \Tr_{\rm query}\bigl((I-P^L)X'(I-P^L)\bigr)\Bigr]F^{L\dagger}.
\end{aligned}
\end{equation}
Let $X^{w}_0=\proj0\otimes\proj\varnothing_L$ and $X^{w}_s(\eta,\beta)=\mathcal C^{w}_{s,\beta_s}\bigl( W_{s-1}X^{w}_{s-1}(\eta,\beta)W_{s-1}^\dagger\bigr)$.

\begin{claim}[Tracing out the right list]\label{appA:images}
For every history $\eta$ of length $i$ and every $\beta\in\{0,1\}^i$,
\[
 \Tr_R\proj{\phi_i(\eta,\beta)}=X^{w}_i(\eta,\beta).
\]
\end{claim}

\begin{proof}
If a birth occurs at a call $s$ with $u_s=0$, both sides are zero: the left side by Claim~\ref{appA:record}, and the right side because $w_s=0$. Otherwise, drop the projectors $\Theta$ using Claim~\ref{appA:support}. Let $s_1<\dots<s_m$ be the births in the first $i$ calls. For each birth, introduce a $d$-dimensional register $Y_a$, and define
\begin{equation}\label{eq:c6-mode-product}
 \Xi_m:\ \ket{y_1}\otimes\cdots\otimes\ket{y_m}\ \longmapsto\
       b^\dagger(v_{s_m},y_m)\cdots b^\dagger(v_{s_1},y_1)
       \ket{\varnothing}_R .
\end{equation}
The map $\Xi_m$ preserves inner products. Indeed, commute each annihilation operator to the right using $b(v,y')b^\dagger(v',y)=\langle v|v'\rangle\delta_{yy'} +b^\dagger(v',y)b(v,y')$. Since the $v_s$ are orthonormal, the inner product of two images is $\prod_a\delta_{y_ay_a'}$.

We construct $\phi_i=(I\otimes\Xi_m)\chi_i$ by induction. Since $\Xi_m$ is an isometry, replacing the right list by $Y_1,\dots,Y_m$ leaves the reduced state unchanged. Initially there are no such registers. A retrieval acts only on the query, left list and memory, and gives the first map in~\eqref{appA:weighted}.

At a birth, let $\chi'=(I-P^L)\pi^{c_i}_iW_{i-1}\chi_{i-1}$. Equation~\eqref{appA:birth-amplitude} and the definition of $\Xi_m$ show that the new right pair stores the incoming query in $Y_m$:
\[
 \chi_i=\sqrt{w_i/d}\,F^L\bigl(\ket{\psi_i}\otimes\chi''\bigr),
\]
where $\chi''$ is $\chi'$ with its query register relabelled as $Y_m$. Tracing out $Y_m$ therefore traces out that incoming query and gives exactly the second map in~\eqref{appA:weighted}, including its correlations with the other registers. Induction proves $\Tr_{Y_1\cdots Y_m}\proj{\chi_i}=X_i^w$, and the isometry $\Xi_m$ then gives the claim.
\end{proof}

Tracing out the history registers removes cross terms between different $\eta$, and tracing out the right list removes cross terms between different $\beta$ by~\eqref{appA:no-cross}. Claim~\ref{appA:images} therefore gives
\begin{equation}\label{appA:weighted-state}
 \Tr_{OLR}\proj{\Psi_5}=\rho^{w},\qquad
 \rho^{w}=\sum_{\eta,\beta}\Tr_L\bigl(W_qX^{w}_q(\eta,\beta)
   W_q^\dagger\bigr).
\end{equation}

\paragraph{Part 3: make the birth weights uniform.}
The weights $w_s$ are at most one because new labels must avoid the span of earlier outcomes. To recover the cache, set all weights to one and call the resulting maps in~\eqref{appA:weighted} $\mathcal C_{s,b}$. Let $X_0=X^{w}_0$ and let $X_s(\eta,\beta)$ follow the same recursion as $X^{w}_s(\eta,\beta)$ with these maps.

This can give weight to histories absent from $\Psi_5$, so we bound the change for every history, including those never reached by the weighted process.

\begin{claim}[The weighted process on all histories]\label{appA:histories}
Let $1\le s\le q$, let $c_1,j_1,\dots,c_{s-1},j_{s-1}$ be arbitrary, and let $X$ be an arbitrary positive operator with a definite number $\nu<d$ of left pairs. Then the following hold. \begin{enumerate}\renewcommand{\labelenumi}{(\roman{enumi})}\setlength\itemsep{0pt}
\item The maps $\mathcal C^{w}_{s,b}$, $\mathcal C_{s,b}$ and $\mathcal C_{s,b}-\mathcal C^{w}_{s,b}$ are completely positive, for every $c_s$, $j_s$ and $b\in\{0,1\}$.
\item $\sum_{c_s,j_s,b}\Tr\bigl[(\mathcal C_{s,b}-\mathcal C^{w}_{s,b})(X)\bigr] \le\frac{s-1}{d}\Tr X$.
\item $\sum_{c_s,j_s,b}\mathcal C_{s,b}(X) =\sum_c\bigl(\mathcal R_{\Delta^c,\nu}+\mathcal B_\nu\bigr) (\pi^c_sX\pi^c_s)$, the call~\eqref{eq:c6-cache} of the label cache with the dephasing $\Delta^c$ in the basis $\psi^c_1,\dots,\psi^c_d$. This map preserves the trace.
\end{enumerate}
Consequently $\bigl\|\sum_{\eta,\beta}\bigl(X_q(\eta,\beta)-X^{w}_q(\eta,\beta)\bigr) \bigr\|_1\le q(q-1)/(2d)$.
\end{claim}

\begin{proof}
For (i), retrieval is unchanged. A birth is a composition of completely positive maps with weight $w_s/d$, and the difference has the same form with weight $(1-w_s)/d$. Both weights are~nonnegative.

For (ii), only births contribute. Since $F^L$ preserves norms, $\Tr[(\mathcal C_{s,1}-\mathcal C^{w}_{s,1})(X)] =\frac{1-w_s}{d}\Tr[(I-P^L)\pi^{c_s}_sX\pi^{c_s}_s]$. For each fixed $c_s$, summing $1-w_s=\bra{\psi_s}\Pi_{\mathcal K_{s-1}}\ket{\psi_s}$ over its basis gives $\dim\mathcal K_{s-1}\le s-1$. The remaining sum over $c_s$ is bounded by $\Tr X$.

For (iii), on lists with $\nu$ left pairs, $P^L=P_\nu$ and $F^{L\dagger}=F_{\nu-1}^\dagger$. The outer $F^L$ is $F_{\nu-1}$ for a retrieval and $F_\nu$ for a birth. Summing retrievals over $j_s$ gives $\mathcal R_{\Delta^c,\nu}$, since $\sum_j\proj{\psi^c_j}\,\cdot\,\proj{\psi^c_j}=\Delta^c$. Summing births gives $\mathcal B_\nu$, since $\sum_j\proj{\psi^c_j}=I$. The two branches have complementary effects $P_\nu$ and $I-P_\nu$, and $\sum_c\pi^c_s=I$, so the sum preserves trace.

To prove the final bound, first note that $X_s(\eta,\beta)\ge X^{w}_s(\eta,\beta)\ge0$ for every history and flag sequence. This holds with equality for $s=0$ and follows by induction from~(i): with $Z=W_{s-1}X_{s-1}W_{s-1}^\dagger$ and $Z^{w}=W_{s-1}X^{w}_{s-1}W_{s-1}^\dagger$, we have
\[
 X_s-X^{w}_s=\mathcal C_{s,\beta_s}\bigl(Z-Z^{w}\bigr)
   +\bigl(\mathcal C_{s,\beta_s}-\mathcal C^{w}_{s,\beta_s}\bigr)(Z^{w}).
\]
The sum $\sum_{\eta,\beta}(X_q-X^{w}_q)$ is positive, so its trace norm is its trace. By (iii), the uniform process has trace one. This trace is therefore $1-\sum_{\eta,\beta}\Tr X^{w}_q$, the trace lost by the weighted process. Parts (ii) and (iii) bound the loss at call $s$ by $(s-1)/d$ times the incoming trace, which is at most one. Summing gives $q(q-1)/(2d)$.
\end{proof}

\paragraph{Error from step 5.}
The number of left pairs is $\nu=\beta_1+\dots+\beta_s$, which starts at zero and increases by one at each birth. This is the counter in Definition~\ref{def:cache}. With $\Delta_I$ for $A$ and $\Delta_I$ or $\Delta_H$ for $B$, Claim~\ref{appA:histories}(iii) identifies the process with the cache: $\sum_{\eta,\beta}\Tr_L(W_qX_q(\eta,\beta)W_q^\dagger) =\rho^{\mathrm{cache}}$. The final bound in Claim~\ref{appA:histories} and \eqref{appA:weighted-state} then give $\TD(\Tr_{OLR}\proj{\Psi_5},\rho^{\mathrm{cache}})\le q(q-1)/(4d)$. Including the cost of replacing $F^R$ by $E^R$, the error in step~5 is at most $q(q-1)/(2d)+q(q-1)/(4d)\le q(q-1)/d=\varepsilon_5$.

\begin{proof}[Proof of Proposition~\ref{prop:cache}]
Set $\varepsilon_{\mathrm{cache}}= \varepsilon_2+\varepsilon_3+\varepsilon_4+\varepsilon_5$. Adding the five error bounds gives
\[
 \TD\bigl(\rho,\rho^{\mathrm{cache}}\bigr)
 \le\varepsilon_{\mathrm{rec}}+\varepsilon_{\mathrm{cache}}.
\]
Since $4q\le d$, every denominator in $\varepsilon_2$ and $\varepsilon_4$ is at least $d/2$. Hence $\varepsilon_2+\varepsilon_4=O(q^{3/2}/\sqrt d)$, $\varepsilon_3=O(q^2/\sqrt d)$, and $\varepsilon_5=O(q^2/d)$. For $q\ge1$ their sum is $O(q^2/\sqrt d)$. The recording term is $\varepsilon_{\mathrm{rec}}=O(q^2/d^{1/8})$, proving the proposition.
\end{proof}

\section{The exact certificate for qubits}\label{sec:certificate}

We verify the support and linear identities used in Section~\ref{sec:qubit-certificate}. All calculations use exact integer or rational arithmetic.

\subsection{Agreement on channel inputs}\label{sec:gauge-proof}

Let $\mathcal W$ be the complex span of the Choi operators of qubit channels. It consists exactly of the operators $H$ for which $\Tr_{A_O}H$ is proportional to $I_{A_I}$. Every channel Choi operator satisfies this condition. Conversely, take Hermitian $H$ with $\Tr_{A_O}H=hI$, and let $J_0=I/2$ be the completely depolarizing Choi operator. Then $H_0=H-hJ_0$ has zero partial trace, and $J_0\pm\epsilon H_0$ are channel Choi operators for sufficiently small $\epsilon>0$. Thus $H$ lies in their linear span. Every operator with partial trace proportional to the identity is a complex linear combination of such Hermitian operators, proving the claim.

For a subsystem $S$, write $ {}_SC=(I_S/\dim S)\otimes\Tr_SC$, with the factors restored to their original order. Define
\[
 \Lambda_A(C)=C-{}_{A_O}C+{}_{A_IA_O}C,\qquad
 \Lambda_B(C)=C-{}_{B_O}C+{}_{B_IB_O}C.
\]
In the Pauli basis $\sigma\otimes\tau$ on $A_I\otimes A_O$, $\Lambda_A$ removes precisely the terms with $\tau=I$ and $\sigma\ne I$. It is therefore the orthogonal projector onto $\mathcal W$ on slot $A$; the same holds for $B$.

The output is obtained by contracting with $J_A\otimes J_B$. An operator gives zero on all channel pairs exactly when it is orthogonal to $\mathcal W\otimes\mathcal W$, with the input and output factors unrestricted. The contraction is bilinear, but this orthogonality statement holds because $\mathcal W$ is closed under complex conjugation. Hence $\mathcal G=\ker(\Lambda_A\Lambda_B)$.

\subsection{The support basis}\label{sec:support-proof}

The rows of $E$ are indexed by $(p,a_i,a_o,b_i,b_o,c,t)\in\{0,1\}^7$. Its columns are listed below; unlisted components vanish, and all free indices take values in $\{0,1\}$.
\begin{center}
\begin{tabular}{@{}lll@{}}
\toprule
column & target path & nonzero components\\
\midrule
$8a_i+4a_o+2b_i+b_o$ & bypasses both boxes & $v_{p,a_i,a_o,b_i,b_o,c,p}=1$\\
$16+2b_i+b_o$ & passes through $A$ & $v_{p,p,t,b_i,b_o,c,t}=1$\\
$20+2a_i+a_o$ & passes through $B$ & $v_{p,a_i,a_o,p,t,c,t}=1$\\
$24$ & undergoes the switch & $v_{p,p,s,s,t,0,t}=v_{p,s,t,p,s,1,t}=1$\\
\bottomrule
\end{tabular}
\end{center}
Under the link~\eqref{eq:vector-link}, these four groups give the operators in~\eqref{eq:four-kraus}, up to a factor $\sqrt2$, multiplied respectively by $U_{a_o,a_i}V_{b_o,b_i}$, $V_{b_o,b_i}$, $U_{a_o,a_i}$, and $1$. Thus $\operatorname{ran}E\subseteq\mathcal V$. An integer left inverse $E^+$ satisfies $E^+E=I_{25}$, proving that these columns are independent.

For the reverse inclusion, choose $U,V$ and a row vector $n$ that annihilates the four vectors in~\eqref{eq:four-kraus}. Every $v\in\mathcal V$ satisfies
\[
 \sum_{p,a_i,a_o,b_i,b_o,c,t}
 n_{p,c,t}\,U_{a_o,a_i}V_{b_o,b_i}\,
 v_{p,a_i,a_o,b_i,b_o,c,t}=0.
\]
We take pairs from matrices of the form
\[
 \begin{pmatrix}g_0+\mathrm ig_3&g_2+\mathrm ig_1\\
 -g_2+\mathrm ig_1&g_0-\mathrm ig_3\end{pmatrix},
 \qquad g_0,\dots,g_3\in\{-1,0,1\}\text{ not all zero}.
\]
They are proportional to unitaries, and rescaling does not change the constraint. Forty pairs give $103$ selected equations with Gaussian-integer coefficient matrix $R\in\mathbb Z[\mathrm i]^{103\times128}$. Stacking $R$ above $E^+$ gives a square matrix $G$ that is invertible modulo $5$ after substituting $\mathrm i\mapsto2$. This substitution is a ring homomorphism because $2^2\equiv-1\pmod5$, so $G$ is also invertible over $\C$. Thus $R$ has rank $103$ and $\dim\mathcal V\le128-103=25$. Together with $\operatorname{ran}E\subseteq\mathcal V$, this proves Lemma~\ref{lem:support}.

\subsection{The linear functional}\label{sec:functional-proof}

The functional in Lemma~\ref{lem:exact-obstruction} is
\begin{align*}
 f(X)={}&X_{24,24}+2X_{0,3}-4X_{0,16}+6X_{0,19}
 -2X_{2,18}-2X_{3,6}\\
 &-2X_{3,19}-2X_{3,24}-2X_{4,7}-2X_{6,19}.
\end{align*}
To verify parts~(a)--(c), substitute $C=EXE^\top$ into the two sets of comb equations~\eqref{eq:ordered-comb} and the gauge equation $\Lambda_A\Lambda_B(C)=0$. Each gives a homogeneous linear system in the entries of the real symmetric matrix $X$. Exact row reduction expresses $f$ as a rational linear combination of the equations in each system. It therefore vanishes on all three solution spaces.

\subsection{The diluted-switch coordinates}\label{sec:diluted-coordinates}

Define the diagonal matrix $X_T$ by
\[
 (X_T)_{kk}=\begin{cases}
 (1-\lambda)(1-\mu)/8,&0\le k\le15,\\
 \lambda(1-\mu)/4,&16\le k\le19,\\
 (1-\lambda)\mu/4,&20\le k\le23,\\
 \lambda\mu/2,&k=24.
 \end{cases}
\]
The first sixteen columns of $E$ are $\sqrt2\,\ket+_c\otimes\vket I_{pt}\otimes \ket{a_i,a_o,b_i,b_o}$, so their contribution is
\[
 (1-\lambda)(1-\mu)\,\proj+_c\otimes\vket I\vbra I_{pt}
 \otimes\tfrac12 I_{A_I}\otimes I_{A_O}
 \otimes\tfrac12 I_{B_I}\otimes I_{B_O}.
\]
This leaves the target unchanged and feeds a maximally mixed dummy state through each slot. The next two groups send the target through $A$ only or $B$ only, with a dummy state in the other slot. The last column is $\sqrt2$ times the switch vector. Hence $EX_TE^\top=C_T$. Since $f$ contains only one diagonal entry, $f(X_T)=X_{24,24}=\lambda\mu/2$, proving part~(d).

The \href{https://github.com/dtlics/Indefinite-Causal-Order/tree/826e646d9d26b51006c83d756d91435bcd171017/proofs/lean/certificates}{supplementary data and verifier}~\cite{repo} supply $E^+$, the selected unitary pairs and null vectors, the inverse of $G$ modulo $5$, and the rational combinations proving the three identities for $f$. The verifier reconstructs the matrices and checks every identity in exact arithmetic.

\bibliographystyle{unsrtnat}
\bibliography{references}
\end{document}